\documentclass[sn-aps,iicol]{sn-jnl}

\usepackage{graphicx}%
\usepackage{multirow}%
\usepackage{amsmath,amssymb,amsfonts}%
\usepackage{amsthm}%
\usepackage{mathrsfs}%
\usepackage[title]{appendix}%
\usepackage{xcolor}%
\usepackage{textcomp}%
\usepackage{manyfoot}%
\usepackage{booktabs}%
\usepackage{algorithm}%
\usepackage{algorithmicx}%
\usepackage{algpseudocode}%
\usepackage{listings}%

\usepackage{multicol}
\usepackage{lineno}

\theoremstyle{thmstyleone}%
\theoremstyle{thmstyletwo}%

\theoremstyle{thmstylethree}%

\begin{document}

\title[Article Title]{Scalar NSI: A unique tool for constraining absolute neutrino masses via neutrino oscillations}


\author[2,1]{\fnm{Abinash} \sur{Medhi}}\email{amedhi0@rnd.iitg.ac.in}
\equalcont{These authors contributed equally to this work.}

\author[1]{\fnm{Arnab} \sur{Sarker}}\email{arnabs@tezu.ac.in}
\equalcont{These authors contributed equally to this work.}

\author*[1]{\fnm{Moon Moon} \sur{Devi}}\email{devimm@tezu.ac.in}
\equalcont{These authors contributed equally to this work.}

\affil*[1]{\orgdiv{Department of Physics}, \orgname{Tezpur University}, \orgaddress{\city{Napaam, Tezpur}, \postcode{784028}, \state{Assam}, \country{India}}}

\affil[2]{\orgdiv{Department of Physics}, \orgname{IIT Guwahati}, \orgaddress{\city{Guwahati}, \postcode{781039}, \state{Assam}, \country{India}}}

\abstract{In the standard interaction scenario, a direct measurement of absolute neutrino masses via neutrino oscillations is not feasible, as the oscillations depend only on the mass-squared differences. However, scalar non-standard interactions (SNSI) can introduce sub-dominant terms in the neutrino oscillation Hamiltonian that can directly affect the neutrino mass matrix, thereby making SNSI a unique tool for neutrino mass measurements. In this work, for the first time, we constrain the absolute masses of neutrinos by probing SNSI. We have explored the constraints on the lightest neutrino mass with different choices of $\delta_{CP}$ and $\theta_{23}$ for both neutrino mass hierarchies. We show that a bound on the neutrino mass can be induced in the presence of SNSI at DUNE. We find that the lightest neutrino mass can be constrained with $\eta_{\tau\tau}$ for normal mass hierarchy irrespective of the octant of $\theta_{23}$ and the value of the CP phase $\delta_{CP}$. This study suggests that SNSI can serve as an interesting avenue to constrain the absolute neutrino masses in long-baseline neutrino experiments via neutrino oscillations.}


\maketitle

\section{Introduction}\label{sec1}
The discovery of neutrino oscillations essentially confirms that neutrinos are massive \cite{Super-Kamiokande:1998kpq,SNO:2002tuh,NOvA:2016kwd,T2K:2013ppw} which provides convincing evidence in favor of physics beyond the Standard Model (BSM). Nonetheless, the absolute mass of neutrinos and the mass ordering are still not precisely known. Neutrino oscillations imply that neutrino flavors mix with each other indicating non-zero masses of neutrinos. However, it cannot provide a direct measurement of the absolute neutrino masses as it depends only on the mass-squared differences. The recent limits on the mass squared splittings can be found in reference \cite{Esteban:2020cvm}. Neutrino masses are represented by $m_{i}$ with orderings $m_{3}>m_{2}>m_{1}$ termed as normal hierarchy (NH) and $m_{2}>m_{1}>m_{3}$ as inverted hierarchy (IH). 

Experiments around the world are in constant pursuit to provide a stringent bound on the neutrino masses \cite{Gariazzo:2018pei,Gerbino:2015ixa,Drexlin:2013lha}. Different sources of data like cosmological \cite{Abazajian:2011dt,Vagnozzi:2017ovm,DiValentino:2021hoh,Tristram:2023haj}, neutrino oscillations \cite{Esteban:2020cvm} and beta-decay \cite{Otten:2008zz,Holzschuh:1992xy} will contribute towards imposing constraints on the neutrino mass \cite{Cao:2021ptu}. A bound on the sum of neutrino masses $\sum m_{i}<$ 0.12 eV (at 95\% CL) is established from cosmology \cite{Planck:2018vyg,Dvorkin:2019jgs,FrancoAbellan:2021hdb,Wong:2011ip,Lesgourgues:2006nd}. However, due to various degeneracies and cosmological assumptions, a stringent limit on the absolute mass of neutrinos is not possible via the cosmological approach. This can be overcome using two different approaches viz. direct and indirect.  In the direct approach, the neutrino mass is estimated by measuring the endpoint of the electron energy spectrum in the $\beta-$decay process. The recent bound on the neutrino mass provided by the KArlsruhe TRItrium Neutrino (KATRIN) experiment \cite{KATRIN:2021dfa} in this approach is $m_\nu<0.45$ eV at 90 \% CL \cite{Katrin:2024tvg,KATRIN:2021uub}. Whereas the neutrinoless double-beta decay ($0\nu\beta\beta$) \cite{Rodejohann:2011mu} is used as an indirect approach to measure the neutrino masses. The mass difference between the parent and daughter nuclei in the $0\nu\beta\beta$ process is used to constrain the neutrino mass. The $0\nu\beta\beta$ decay is only possible because neutrinos are Majorana particles and it can provide a bound on the neutrino mass via $m_{\beta\beta}$ = $\sum U_{ei}^2m_{i}$ and its current bound is $m_{\beta\beta}$ $<$ 75 - 180 meV \cite{GERDA:2020xhi,CUORE:2019yfd,KamLAND-Zen:2016pfg}.
The mass of neutrinos is five orders of magnitude smaller than any other Standard Model fermions. The determination of neutrino mass and its underlying mechanism will shed light on various fundamental unknowns of particle physics. 

The challenges in measuring the absolute neutrino masses can arise from various reasons. The weak interaction of neutrinos with matter makes it challenging to measure the neutrino mass with high precision. The background noises are generally very high in such mass measurement experiments. Also, the complexity of the neutrino experiments leads to higher systematic errors. Despite these challenges, a significant development can be seen in the neutrino mass bounds. In recent years, there have been several important results, such as the first direct measurement of the neutrino mass squared difference and the most stringent limits on the $0\nu\beta\beta$ decay half-life \cite{Singh:2023ked}.

The neutrino oscillation experiments can probe the mass-squared differences of the neutrinos but are not sensitive to the absolute masses. However, the presence of scalar non-standard interactions (SNSIs) can bring a direct dependence of absolute neutrino masses via neutrino oscillations \cite{Ge:2018uhz,Smirnov:2019cae, Yang:2018yvk, Khan:2019jvr, Babu:2019iml, Gupta:2023wct,Singha:2023set,Denton:2022pxt,Dutta:2022fdt,ESSnuSB:2023lbg}. Neutrinos coupling via a scalar is interesting as it affects the neutrino mass term which can be further probed to constrain the absolute masses of neutrinos. In our earlier studies \cite{Medhi:2021wxj,Medhi:2022qmu,Sarker:2023qzp,Sarker:2024ytu}, we found that the SNSI can significantly impact the physics sensitivities of long baseline (LBL) experiments. 
The presence of SNSI can provide a unique pathway to explore the absolute neutrino masses. A stringent bound on the neutrino mass will not only provide a better understanding of the properties of neutrinos but also help in explaining the physics BSM. 

In this work, for the first time, we obtain a constraint on the lightest neutrino mass ($m_\ell$, where $\ell$=1 for NH and $\ell$=3 for IH) by probing neutrino oscillation experiments. We have performed the study in a model-independent way. As the SNSI contribution scales linearly with matter density, it makes long-baseline experiments a suitable candidate to explore its effects. We consider a flavor-conserving texture of SNSI. We probe the impact of SNSI considering one element at a time, at the upcoming long baseline experiment DUNE \cite{DUNE:2015lol,DUNE:2016hlj,DUNE:2020txw,DUNE:2021cuw}.  We constrain the lightest neutrino mass considering that $\eta_{\alpha\beta}$ is well restricted from other experiments. We point out that the SNSI elements can constrain the neutrino mass with a marginally better constraining capability for NH than that of IH. We also note a slight variation in the constraining capability at DUNE with respect to different choices of SNSI elements. We also checked how the constraining capability of the experiment changes with different values of $\delta_{CP}$ and $\theta_{23}$. Interestingly we found the constraining capability of the experiment remains intact for all of these scenarios. The findings of our analysis can help in estimating the absolute neutrino masses in neutrino oscillation experiments and will have the potential to play a key role in our quest to understand the nature of neutrinos.

We have organized the paper as follows, in section \ref{sec:formalism}, we provide a detailed formalism of SNSI. Then in section \ref{sec:meth}, we discuss the methodology used in our study, the experimental details of DUNE and the statistical framework adopted for this work. In section \ref{sec:probability}, we explore the impact of the SNSI elements at the probability level. We also provide an approximate analytical expression of probability in section \ref{sec:analytical_Pmue}. In section \ref{sec:Pmue} and \ref{sec:Pmue_space}, we have explored the dependence of neutrino oscillation probabilities on the neutrino mass and its impact in the $\delta_{CP}$ parameter space respectively.  We have put bounds on the lightest neutrino mass for NH and IH in section \ref{sec:bound_m}. The correlation in ($\eta_{\alpha\alpha}$--$m_{\ell}$) parameter space is explored in section \ref{sec:correlation_m}. Finally, we conclude our findings in section \ref{sec:conclusion}.

\section{Scalar NSI Formalism} \label{sec:formalism}
The interaction of neutrinos with matter occurs via the mediators $W^{\pm}$, $Z^{0}$ and it impacts the neutrino oscillations \cite{Pontecorvo:1957cp,Pontecorvo:1957qd,Pontecorvo:1967fh,ParticleDataGroup:2020ssz}. These matter effects appear as an additional potential term in the Hamiltonian for neutrino oscillations \cite{Wolfenstein:1977ue, Nieves:2003in, Nishi:2004st, Maki:1962mu}. The scalar interactions of neutrinos are an interesting possibility as neutrinos can couple with a scalar (Higgs boson) with non-zero vacuum expectation to generate its mass.

\vspace{-10pt}
\begin{figure}[!h]
    \centering
    \includegraphics[width=0.5\linewidth, height =4cm]{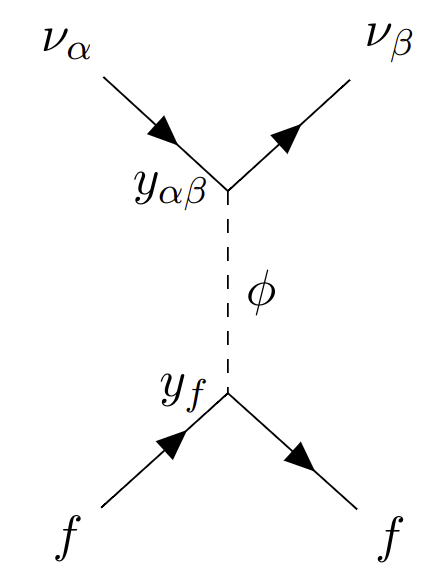}
    \caption{A Feynman diagram representing the coupling of neutrinos via a scalar mediator $\phi$.}
    \label{fig:enter-label}
\end{figure}

\noindent The Lagrangian for such non-standard coupling of $\nu$'s with the environmental fermions via a scalar mediator can be written as \cite{Ge:2018uhz, Babu:2019iml,Medhi:2021wxj,Medhi:2022qmu},
\begin{equation}\label{eq:Lag_SNSI}
    \mathcal{L}_{SNSI}=\frac{y_f y_{\alpha\beta}}{m_\phi^2}~\left[\overline{\nu_{\alpha}}(p_{3})\nu_{\beta}(p_{2})\right]\left[\overline{f}(p_{1})f(p_{4})\right]
\end{equation}
where,
\begin{itemize} 
\item $\alpha, \beta = e, \mu, \tau$ are the neutrino flavors.
\item $f$ ($\bar f)$ is matter fermion (anti-fermion).
\item $y_{\alpha \beta}$ is the Yukawa couplings of neutrinos with the scalar mediator $\phi$.
\item $y_f$ is the Yukawa coupling of the scalar mediator with the environmental fermions $f$
\item $m_\phi$ is the mass of the scalar mediator $\phi$. 
\end{itemize}
\noindent The modification in the Lagrangian $\mathcal{L}$ leads to a correction in the Dirac equation as seen below,
\begin{equation}\label{eq:Mod_Dirac_eq}
\overline{\nu}_{\beta}\left[i\partial_{\mu}\gamma^{\mu}+\left(M_{\beta\alpha}+\frac{\sum_{f}n_{f}y_{f}Y_{\alpha\beta}}{m_{\phi}^{2}}\right)\right]\nu_{\alpha}=0
\end{equation}

\noindent The quantity $n_f$ defines the number density of environmental fermions which is assumed to be non-relativistic considering the Earth matter. The effect of SNSI is inversely proportional to the square of the mediator mass. Hence, a light mediator can produce a notable impact on the SNSI terms. So far there is no stringent bound on the mediator mass for such an ultralight mediator. However, in reference \cite{Cordero:2022fwb}, the authors considered the mediator mass to be approximately $10^{-22}$ eV/$c^2$ and studied its impact on the upcoming European Spallation Source neutrino Super Beam (ESS$\nu$SB) experiment \cite{ESSnuSB:2021azq}. The mass of the scalar field, inferred from the anisotropies of the Cosmic Microwave Background (CMB), falls within the range of $m_\phi = 10^{-24}$ to $10^{-22}$ eV/$c^2$ \cite{Urena-Lopez:2019kud}. Furthermore, based on observations of galaxy rotation curves, the mass of ultra-light dark matter is estimated to be in the range of $m_\phi = 0.5 \times 10^{-23}$ to $10^{-21}$ eV/$c^2$ \cite{Bernal:2017oih, Urena-Lopez:2017tob}. An updated constraint on the SNSI parameters from the solar neutrino sector is calculated in reference \cite{Denton:2024upc}. As shown in reference \cite{Ge:2018uhz, Babu:2019iml}, the SNSI can appear as a correction to the neutrino mass matrix. The effective Hamiltonian in the presence of SNSI can be framed as,
\begin{equation}\label{eq:Hamil}
    \mathcal{H}\thickapprox E_{\nu}+\frac{M_{eff}M_{eff}^{\dagger}}{2E_{\nu}}\pm V_{SI}
\end{equation}
where, $M_{eff}\equiv M+M_{SNSI}$ with $M_{SNSI}\equiv \sum_{f}n_{f}y_{f}y_{\alpha\beta}/m_{\phi}^{2}$. The neutrino mass matrix can be diagonalized by a modified mixing matrix as $\mathcal{U}^{'}$ = $P\mathcal{U}Q^{\dagger}$, where the Majorana rephasing matrix Q can be absorbed by $QD_{\nu}Q^{\dagger}=D_{\nu}=diag(m_{1},m_{2},m_{3})$. The unphysical diagonal rephasing matrix, P can be rotated away into the SNSI contribution as follows,
\begin{equation}
M_{eff}=\mathcal{U}D_{\nu}\mathcal{U}^{\dagger}+P^{\dagger}M_{SNSI}P=M+ \delta M    
\end{equation}
In the standard case, a common $m_{i}^{2}$ term can be subtracted to obtain a dependence of neutrino oscillations on the mass-squared splittings i.e. $\Delta m_{21}^{2}$ and $\Delta m_{31}^{2}$. However, in the presence of SNSI, the cross terms i.e. $M \delta M^{\dagger}$ and $M^{\dagger}\delta M$ bring a direct dependence on the absolute masses of neutrinos as no common term can be subtracted from the mass matrix. Hence, oscillation probabilities will also directly depend on the absolute mass of neutrinos. We parametrize the SNSI contribution $\delta M$ in a model-independent way following previous studies as \cite{Ge:2018uhz,Medhi:2021wxj,Medhi:2022qmu},
\begin{equation}\label{eq:DM_parameterization}
\delta M\equiv S_{m}\left(\begin{array}{ccc}
\eta_{ee} & \eta_{e\mu} & \eta_{e\tau}\\
\eta_{e \mu}^* & \eta_{\mu\mu} & \eta_{\mu\tau}\\
\eta_{e \tau}^* & \eta_{\mu\tau}^* & \eta_{\tau\tau}
\end{array}\right)
\end{equation}
where, the dimensionless parameter $\eta_{\alpha\beta}$ quantifies the strength of SNSI and $S_{m}$ is used as a rescaling factor to express $\delta M$ in mass dimension. We have fixed the value of the rescaling factor $S_{m}$ to $\sqrt{2.55\times10^{-3}}$ eV throughout the analysis. This corresponds to a typical value of atmospheric mass squared splitting. As SNSI scales linearly with matter density, the $\eta_{\alpha\beta}$ parameters can be scaled as,
\begin{equation}
    \eta_{\alpha\beta} = \eta_{\alpha\beta}^{\rm (true)} \left(\frac{\rho_{\text  {\tiny expt}} - \rho_0}{\rho_0}\right)\;
\end{equation}
Here, $\eta_{\alpha\beta}^{\rm (true)}$ is the true value of the SNSI parameter, $\rho_{\text{\tiny expt}}$ is the average matter density experienced by neutrinos in the chosen LBL experiment, and $\rho_0$ is the average matter density. In our simulations, all the $\eta_{\alpha\beta}$ values are defined at an average matter density $\rho_0=2.9 ~gm/cm^{3}$ even though the impact of rescaling $\eta_{\alpha\beta}$ is nominal in long baseline sector. We can frame the Lagrangian for a single non-zero $\eta_{\alpha\alpha}$ element as,
\begin{equation}\label{eq:Lag_SNSI}
    \mathcal{L}_{SNSI}=\frac{y_f y_{\alpha\alpha}}{m_\phi^2}\left[\overline{\nu_{\alpha}}(p_{3})\nu_{\alpha}(p_{2})\right]\left[\overline{f}(p_{1})f(p_{4})\right]
\end{equation}
\vspace{-10pt}

\begin{table*}[!t]
    \centering
    \begin{tabular}{|c|c|c|}
    \hline 
    \rule{0pt}{12pt} Parameter & Best-fit Values & Marginalization Range \tabularnewline
    \hline 
    \hline 
    \rule{0pt}{12pt} $\theta_{12}$ & $34.51^{\circ}$ & fixed\tabularnewline
    \rule{0pt}{12pt} $\theta_{13}$ & $8.44^{\circ}$ & fixed\tabularnewline
    \rule{0pt}{12pt} $\theta_{23}$ & $LO:43^{\circ}$, $HO:47^{\circ}$ & $[40^{\circ},50^{\circ}]$\tabularnewline
    \rule{0pt}{12pt} $\delta_{CP}$ & $-\pi/2$ & fixed\tabularnewline
    \rule{0pt}{12pt} $\Delta m_{21}^{2}$ & $7.56\times10^{-5}eV^{2}$ & fixed\tabularnewline
    \rule{0pt}{12pt} $\Delta m_{31}^{2}$ (NH)& $2.55\times10^{-3}eV^{2}$ & $[2.25,2.65]\times10^{-3}eV^{2}$\tabularnewline
    \rule{0pt}{12pt} $\Delta m_{31}^{2}$ (IH)& $-2.497\times10^{-3}eV^{2}$ & $[-2.65,-2.25]\times10^{-3}eV^{2}$ \tabularnewline
    \hline 
    \end{tabular}\\
    \vspace{0.1cm}
    \caption{The oscillation parameter values used in our analysis \cite{NuFIT5.0}, along with the corresponding marginalization ranges.}
    \label{tab:param_val}
\end{table*}

\begin{figure}[!h]
    \centering
    \includegraphics[width=0.5\linewidth, height =4cm]{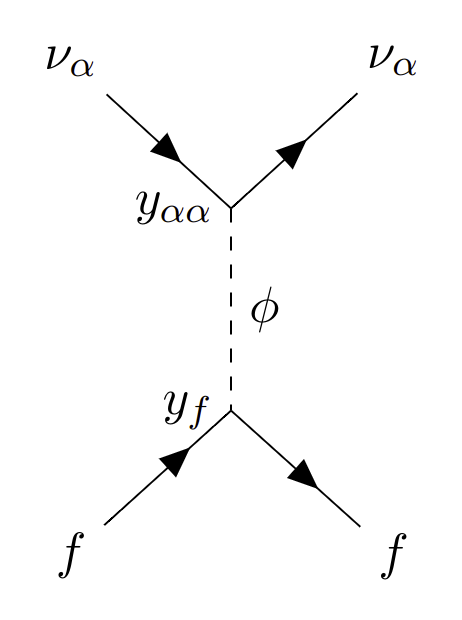}
    \caption{The Feynman diagram for one non-zero diagonal SNSI element, $\eta_{\alpha\alpha}$.}
    \label{fig:feyn2}
\end{figure}

\noindent The Feynman diagram for a single non-zero diagonal SNSI ($\eta_{\alpha\alpha}$) case is shown in Fig. \ref{fig:feyn2}. In this study, we do not take into account any diagrams or cases that involve flavor violation. Consequently, we set the off-diagonal entries to zero as we focus exclusively on the flavour-conserving scenario.

We consider three such cases where the diagonal SNSI elements $\eta_{ee}$, $\eta_{\mu\mu}$ and $\eta_{\tau\tau}$ are non-zero. We explore the direct dependence of neutrino oscillation probabilities on the neutrino mass to put bounds on the lightest neutrino mass. We test with a non-zero diagonal $\eta_{\alpha\alpha}$ element, one at a time. The $M_{eff}$ for such a case when $\eta_{\alpha\alpha} \neq 0$ is shown for reference below, where the explicit dependence on the absolute neutrino masses is illustrated.
\begin{subequations}
\begin{align}
M_{eff} & = \mathcal{U}\cdot D_{\nu}\cdot\mathcal{U}^\dag + S_{m}\rm diag \left( \eta_{ee}, 0, 0\right)\\
M_{eff} & = \mathcal{U}\cdot D_{\nu}\cdot\mathcal{U}^\dag + S_{m}\rm diag \left( 0, \eta_{\mu\mu}, 0\right)\\
M_{eff} & = \mathcal{U}\cdot D_{\nu}\cdot\mathcal{U}^\dag + S_{m}\rm diag \left( 0, 0, \eta_{\tau\tau}\right)
 \label{Meff}
\end{align}
\end{subequations}

We explore the SNSI effects in the long baseline sector taking DUNE as a case study. In section \ref{sec:meth}, we describe the technical details of DUNE and the simulation methodology followed throughout this analysis.

\section{Methodology} \label{sec:meth}
 The long-baseline neutrino experiments will help to explore new physics effects and lead to better constraints of the neutrino mixing parameters. As SNSI scales linearly with matter density, the long-baseline experiments are suitable for probing its effects. This will in turn help in constraining the absolute mass of neutrinos by putting bounds on the SNSI parameters. The possibility of constraining the absolute neutrino masses by probing the effects of SNSI is investigated in this work. From the existing cosmological bound $\sum m_{i}<0.12$ eV (95$\%$ CL) \cite{Planck:2018vyg,Dvorkin:2019jgs,FrancoAbellan:2021hdb,Wong:2011ip,Lesgourgues:2006nd}, we obtain the upper limits of the lightest neutrino masses for both hierarchies. The upper limits of the lightest neutrino masses $m_1$ (NH) and $m_3$ (IH) are 0.03 eV and 0.015 eV following the cosmological bound (as shown in Table \ref{tab:dm_splitting_NO_IO} in Appendix \ref{app:mass_range}). The benchmark values of neutrino oscillation parameters used throughout the analysis are tabulated in table \ref{tab:param_val}.

\subsection{DUNE} 
The Deep Underground Neutrino Experiment (DUNE) \cite{DUNE:2015lol,DUNE:2016hlj,DUNE:2020txw,DUNE:2021cuw} is a next-generation long baseline experiment with a 40 kton Liquid Argon Time Projection Chamber (LArTPC) detector to be placed at the Homestake Mine in South Dakota. The neutrinos for the experiment will be produced at Fermilab's Long-Baseline Neutrino Facility (LBNF), which has a 1.2 MW proton beam capable of delivering $1.1 \times 10 ^{21}$ proton on target (POT) annually. DUNE will have a baseline of 1300km with the near and far detector situated at a distance of 574m and 1300km respectively. The main goals of the experiment are the precise measurement of neutrino oscillation parameters, probing of CP-violation, and identifying the true mass hierarchy of neutrinos. The experiment plans to start data-taking by the end of this decade. In this work, we have used the inputs from the technical design report (TDR) of DUNE to simulate the experiment \cite{DUNE:2021cuw, DUNE:2020ypp}. The detector details and the systematic uncertainties for DUNE are summarized in the table \ref{tab:sys_unc}.

This experiment can be a good probe for exploring the effects of SNSI which scales linearly with the environmental matter density. In our study, we have used the General Long Baseline Experiment Simulator (GLoBES) package \cite{Huber:2004ka,HUBER2007439} to calculate the numerical oscillation probabilities. We have considered the detector details and systematics for the DUNE experiment. We have considered a total runtime of 7 years (3.5$\nu$ + 3.5$\bar{\nu}$) for our simulation studies.  The details of the statistical framework for exploring the physics sensitivities are described in section \ref{sec:chi2}. 

\begin{table}[!h]
    \centering
    \renewcommand{\arraystretch}{1.2}
    \begin{tabular}{c}
    \hline 
    \rule{0pt}{12pt}Experimental Details \tabularnewline
    \hline 
    \hline 
    \rule{0pt}{12pt}DUNE (Baseline = 1300 km) \tabularnewline
    \rule{0pt}{12pt}Runtime = 3.5 yr $\nu + 3.5$ yr $\bar \nu$  \tabularnewline
    \rule{0pt}{12pt}L/E = 1543 km/GeV , Fiducial mass = 40 kt (LArTPC)\tabularnewline
    \end{tabular}

    \begin{tabular}{c|c|c} 
    \hline
    \rule{0pt}{10pt}Detection Channels&Signal Error&Background Error\tabularnewline
    \hline 
    \hline 
    \rule{0pt}{10pt} $\nu_{e}(\bar{\nu_{e}})$ appearance & 2$\%$(2$\%$) & 5$\%$(5$\%$)\tabularnewline
    \rule{0pt}{10pt} $\nu_{\mu}(\bar{\nu_{\mu}})$ disappearance & 5$\%$(5$\%$) & 5$\%$(5$\%$)\tabularnewline
    \rule{0pt}{2pt} & & \tabularnewline
    \hline 
    \end{tabular}
    \vspace{0.1cm}
    \caption{The detector details and the systematic uncertainties for DUNE.\cite{DUNE:2021cuw,DUNE:2020ypp}}
    \label{tab:sys_unc}
\end{table}

\noindent In order to estimate a ballpark range of $\eta_{\alpha\alpha}$ and $m_\ell$ values, we define a quantity $\mathcal{R}_{SNSI}$ as
\begin{equation}
\mathcal{R}_{SNSI} = \Big|\frac{H_{SNSI}\big[i,j\big]}{H_{SI}\big[i,j\big]}\Big|_{max},
\end{equation}
where $H_{SNSI}\big[i,j\big]$ is the $ij^{th}$ element of SNSI contribution to the effective Hamiltonian and $H_{SI}\big[i,j\big]$ is the $ij^{th}$ element of standard Hamiltonian. In addition, we have subtracted the $[1,1]$ element from all the diagonal ratios to remove an overall constant phase. This leaves us with eight ratios out of which we consider the ratio with the maximum value. In Fig. \ref{fig:SNSI_ratio} (Appendix \ref{sec:ratio_sec}), we have explored the parameter space of ($\eta_{\alpha\alpha}$-$m_\ell$) for representative purposes. Depending on the observed contours, we have restricted our choice of $\eta_{\alpha\alpha}$ such that $\mathcal{R}_{SNSI}$ lies below 0.2 (20\%). In our analysis, we have used the values of $\eta_{\alpha\alpha}=\pm0.03$ and $m_\ell=$0 eV, 0.005 eV as a case study. The range of values of SNSI parameters can be explored across different fields, including solar, reactor and accelerator sectors. Notably, in the work \cite{Denton:2024upc}, the authors have put constraints on the SNSI parameters using solar and reactor neutrino data. The choices of SNSI parameters used in our analysis are consistent with the constraints derived from these studies \cite{Dutta:2024hqq, Denton:2024upc}.

\onecolumn
\begin{multicols}{2}
\subsection{Statistical $\chi^{2}$-framework} \label{sec:chi2}
We present our results for constraining the lightest neutrino mass for both normal and inverted hierarchies in the presence of SNSI taking DUNE as a case study. We place bounds on the neutrino masses by taking one diagonal SNSI element $(\eta_{\alpha\alpha})$ at a time. We probe $\eta_{\alpha\alpha}$ on the appearance channel ($P_{\mu e}$) for different choices of $m_\ell$. We particularly focus only on the SNSI parameters $\eta_{ee}$, $\eta_{\mu\mu}$ and $\eta_{\tau\tau}$ and examine DUNE's capability towards constraining the neutrino masses. We further study the correlation of the lightest neutrino mass with the SNSI parameters. We use the same values of neutrino oscillation parameters mentioned above in section \ref{sec:meth}. In order to constrain the lightest neutrino mass, we define a statistical $\chi^{2}$ which is a measure of sensitivity as,
\begin{equation}
\label{eq:chisq}
\small
\chi_{pull}^{2}=\underset{\zeta_{j}}{min}\left(\min_{\eta}  \sum_{i} \sum_{j}
\frac{\left[N_{true}^{i,j} - N_{test}^{i,j} \right]^2 }{N_{true}^{i,j}}+\sum_{i=1}^{k}\frac{\zeta_{i}^{2}}{\sigma_{\zeta_{i}}^{2}}\right)
\end{equation}
\noindent where, $N_{true}^{i,j}$ and $N_{test}^{i,j}$ represents the number of true and test events in the $\{i,j\}$-th bin respectively. Using the pull method described in \cite{Huber:2004ka,Fogli:2002pt}, we incorporate the systematic errors as additional parameters known as nuisance parameters ($\zeta_{k}$) with the systematical errors ($\sigma_{\zeta_{k}}^{2}$). We define the sensitivity as $\Delta\chi^{2}$ to constrain the lightest neutrino masses as shown in Eq. \ref{eq:delchi2}. We have marginalized over all the systematic uncertainties. It is important to note that the mixing parameter values used in the analysis are taken from table \ref{tab:param_val} unless stated otherwise.

\section{Analysis at Probability level}\label{sec:probability}
We have explored the analytic probability expressions for the appearance channel in the presence of diagonal SNSI elements $\eta_{\alpha\alpha}$ in section \ref{sec:analytical_Pmue}. The impact of SNSI on the appearance probability channel for normal and inverted hierarchy considering different cases of lightest neutrino mass is explored using numerically calculated oscillation probabilities in section \ref{sec:Pmue}. In addition, we have also explored the ($\delta_{CP}$ - $m_\ell$) parameter space in the presence of SNSI in section \ref{sec:Pmue_space}.

\subsection{Approximate probability expression}\label{sec:analytical_Pmue}
Approximated expressions of neutrino oscillation probabilities can offer insights into the underlying parameter dependence such as the explicit dependence on the mixing parameters, baseline and neutrino energy. It can also help understand possible degeneracies and optimize experiments for precise measurements of relevant parameters. These analytic expressions in the presence of BSM scenarios can help explore the intricate dependence on the model parameters. A comparison of different analytic expressions is shown in reference \cite{Barenboim:2019pfp}. A detailed analysis of the oscillation probabilities with higher-order contributions in the presence of SNSI is provided in the recent work \cite{Bezboruah:2024yhk}. This offers a comprehensive treatment of oscillation probabilities in the presence of diagonal SNSI elements. These expressions can provide insights into the dependence of oscillation probabilities on the diagonal elements. It also highlights the role of SNSI in modifying the 3$\nu$ standard probabilities. 

The probability contribution of SNSI elements $\eta_{\alpha\alpha}$ to the standard oscillation probabilities can be expressed as shown in Eqs. \ref{eq:pme_etaee}-\ref{eq:pme_etatt}. The probabilities are perturbatively calculated by expanding over the parameters $s_{13}\approx O(\lambda)$, $\alpha\approx O(\lambda^2)$ and $\eta_{\alpha\alpha}\approx O(\lambda^2)$. Here, we have only used the probability expressions as explicitly derived in \cite{Bezboruah:2024yhk} considering terms up to $O(\lambda^4)$. We have expressed only the SNSI contribution to the SI oscillation probabilities. The presence of only one non-zero diagonal element is considered. The probability contribution for $\eta_{ee}$, $\eta_{\mu\mu}$ and $\eta_{\tau\tau}$ up to $O(\lambda^4)$ are presented in Eq. \ref{eq:pme_etaee}, \ref{eq:pme_etamm} and \ref{eq:pme_etatt}, respectively as derived in \cite{Bezboruah:2024yhk}.
\end{multicols}

\begin{subequations}
\begin{align}\label{eq:pme_etaee}
    P_{\mu e}^{(\eta_{ee})} = & \, \Bigg[ 2 s_{13} \sin \left(2 \theta _{12}\right) \sin \left(2 \theta _{23}\right) \cos \left(\delta _{\text{CP}}+\Delta \right) \frac{\sin \left[ \left(A_e-1\right) \Delta\right]}{A_e-1} \frac{\sin \left[ A_e \Delta \right]}{A_e} (m_2-m_1) \nonumber \\
    & + 2  \alpha \, c_{23}^2 \sin^2 \left(2 \theta _{12}\right) \frac{\sin^2 \left[A_e \Delta \right]}{A_e^2} \Big( m_2-m_1 \Big) \Bigg. + 8 s_{13}^2 s_{23}^2 \frac{\sin \left[ \left(A_e-1\right) \Delta \right]}{A_e-1} \Bigg( \frac{\sin \left[  \left(A_e-1\right) \Delta \right]}{A_e-1} \, m_3 \Bigg.  \nonumber\\
    &\quad + \frac{1}{A_e-1} \left[ 2 \, \Delta  \cos \left[ \left(A_e-1\right)\Delta \right]- \left(A_e+1\right) \frac{ \sin \left[ \left(A_e-1\right) \Delta \right]}{A_e-1} \right] \Bigg. \Bigg.\times \Big( m_1 c_{12}^2 +m_2 s_{12}^2 \Big) \Bigg) \Bigg]  \left( \frac{S_m}{\Delta m^2_{31}} \right) \eta_{ee}.
\end{align}
\begin{align}\label{eq:pme_etamm}
P_{\mu e}^{(\eta_{\mu\mu})} = & \Biggr[ s_{13} \sin \left(2 \theta _{12}\right) \sin \left(2 \theta _{23}\right) \frac{\sin \left[ \left(A_e-1\right) \Delta \right]}{A_e-1} \Biggr[  \Bigg( \cos \left(\delta _{\text{CP}}\right) \frac{\sin \left[ \left(A_e-1\right) \Delta \right]}{A_e-1} \Bigg.\Biggr. \nonumber \\
        & \; + \frac{1}{A_e-1} \bigg[ \sin (\Delta ) \cos \left(2 \theta _{23}\right) \cos \left(\Delta  A_e+\delta _{\text{CP}}\right)+\sin \left(\Delta  A_e\right) \cos \left(\delta _{\text{CP}}+\Delta \right) \bigg. \nonumber\\
        & \Biggr. \Bigg. \bigg. \qquad \quad -2 c_{23}^2 \cos \left(\delta _{\text{CP}}+\Delta \right) \frac{\sin \left[ A_e \Delta \right]}{A_e} \bigg] \Bigg) \big(m_2 -m_1 \big) \nonumber \\ 
        & + 4 \, s_{13}^2 \, s_{23}^2 \frac{\sin \left[ \left(A_e-1\right) \Delta \right]}{A_e-1} \Bigg( \cos \left(2 \theta _{12}\right) \frac{\sin \left[\left(A_e-1\right) \Delta  \right]}{A_e-1} \Big( m_2-m_1 \Big) \Bigg. \Biggr.  \nonumber\\
        & +\frac{1}{A_e-1} \bigg[ 2 c_{23}^2 \cos (\Delta )\frac{\sin \left[ A_e \Delta \right]}{A_e} - \Big( \cos \left(2 \theta _{23}\right) \sin (\Delta ) \cos \left[ A_e \Delta \right]+\cos (\Delta ) \sin \left[A_e \Delta  \right] \Big)\bigg] \Big( m_1+m_2 \Big) \nonumber \\
        & \Bigg.  + \frac{2}{A_e-1} \bigg[ \Big(A_e-\cos \left(2 \theta _{23}\right)\Big) \frac{\sin \left[ \left(A_e-1\right) \Delta \right]}{A_e-1} - 2 \Delta  s_{23}^2 \cos \left[ \left(A_e-1\right) \Delta \right]\bigg] m_3 \! \Bigg) \nonumber \\
        & + 2 \,\alpha \, c_{23}^2 \sin ^2\left(2 \theta _{12}\right) \frac{\sin \left[ A_e \Delta \right]}{A_e} \frac{1}{A_e-1} \Bigg( -c_{23}^2 \frac{\sin \left[ A_e \Delta \right]}{A_e}  \Bigg. + \Big[ \sin \left[ A_e \Delta \right]-s_{23}^2 \sin (\Delta ) \cos \left[\left(A_e-1\right) \Delta  \right] \Big]\Bigg)  \nonumber \\
        & \Big( m_2-m_1 \Big)\Biggr] \left(\frac{S_m}{\Delta m^2_{31}}\right) \eta _{\mu \mu }
\end{align}

\begin{align}\label{eq:pme_etatt}
    P_{\mu e}^{(\eta_{\tau\tau})}= & \, \Biggr[ 2 s_{13} s_{23}^2 \sin \left(2 \theta _{12}\right) \sin \left(2 \theta _{23}\right) \frac{\sin \left[\left(A_e-1\right) \Delta  \right]}{A_e-1} \frac{1}{A_e-1} \nonumber\\
    & \; \times \Bigg(\cos \left(\Delta  A_e+\delta _{\text{CP}}\right) \sin (\Delta ) - \cos \left(\delta _{\text{CP}}+\Delta \right) \frac{\sin \left[ A_e \Delta \right]}{A_e} \Bigg) \Big(m_2-m_1\Big) \nonumber \\
    &+ \frac{\sin ^2\left(2 \theta _{23}\right)}{A_e-1} \Biggr[ \alpha  \sin ^2\left(2 \theta _{12}\right) \frac{\sin \left[ A_e \Delta \right]}{4 A_e}  \Biggr. \times \Bigg( \Big[\sin \left[ A_e \Delta \right]-\sin \left[ \left(A_e-2\right) \Delta \right] \Big] -2\frac{\sin \left[ A_e \Delta \right]}{A_e} \Bigg) \Big(m_2 -m_1\Big)\nonumber\\
    & \; -2 s_{13}^2 \frac{\sin \left[\left(A_e-1\right) \Delta  \right]}{A_e-1} \Bigg( \bigg[\cos (\Delta )  \frac{\sin \left[ A_e \Delta \right]}{A_e}-\sin (\Delta ) \cos \left[ A_e \Delta \right] \bigg] \Big( m_1 + m_2 \Big) \Bigg. \nonumber\\
    & \Biggr. \Bigg. \quad \quad + 2 \bigg[ \Delta  \cos \left[ \left(A_e-1\right) \Delta \right]-\frac{\sin \left[\left(A_e-1\right) \Delta \right]}{A_e-1} \bigg] m_3 \Bigg) \Biggr] \Biggr]\times \left(\frac{S_m}{\Delta m^2_{31}}\right) \eta _{\tau \tau }
\end{align}
\end{subequations}
where, 
\begin{equation}
    \alpha = \frac{\Delta m_{21}^2}{\Delta m_{31}^2}, \qquad A_e =\frac{2 E_\nu V_{SI}}{\Delta m_{31}^2}, \qquad \Delta =\frac{\Delta m_{31}^2 L}{4 E_\nu}.
\end{equation}

\begin{figure*}[!b]
\centering
\includegraphics[height=5cm,width=0.32\linewidth]{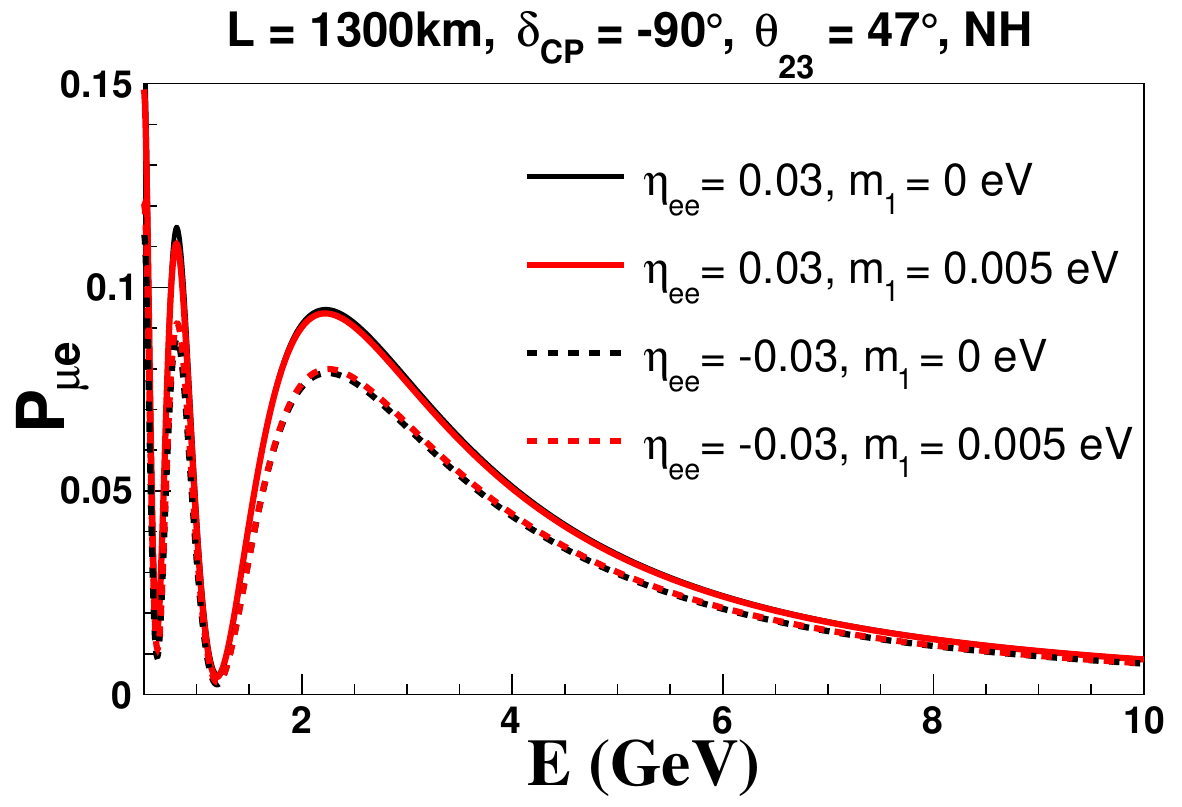}
\includegraphics[height=5cm,width=0.32\linewidth]{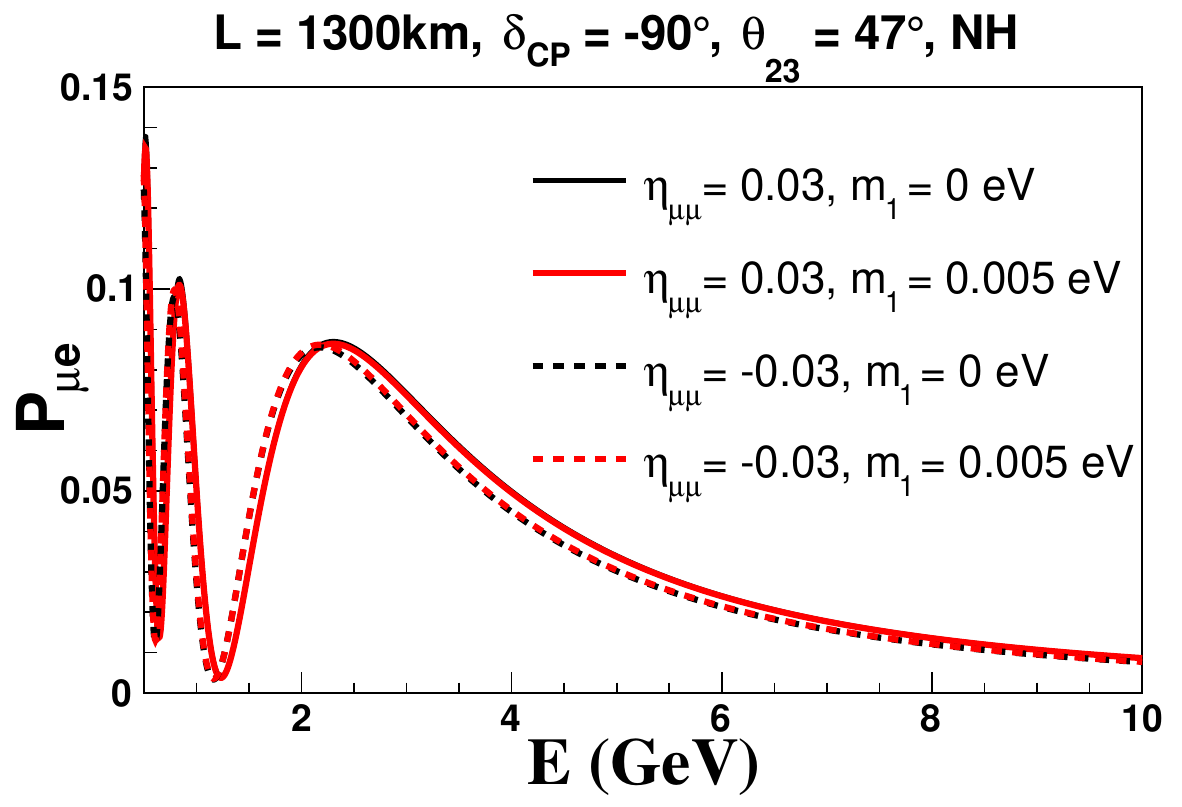}
\includegraphics[height=5cm,width=0.32\linewidth]{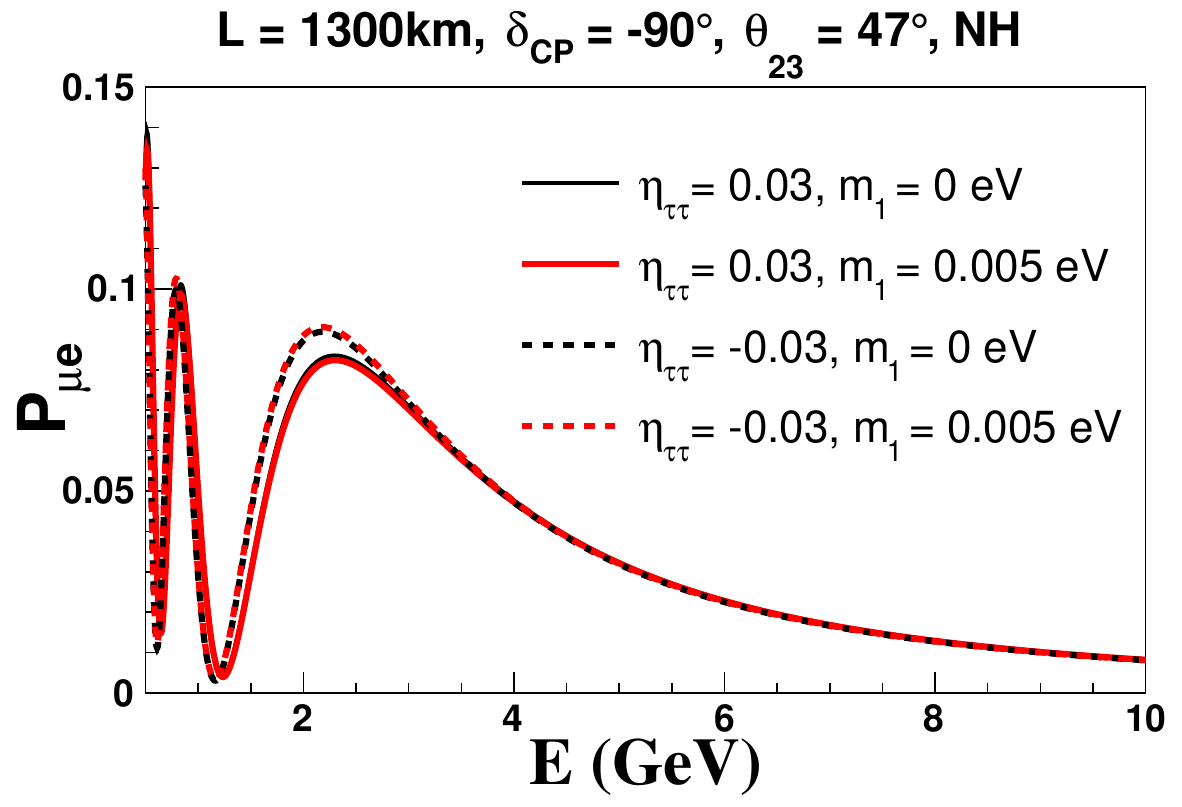}
\includegraphics[height=5cm,width=0.32\linewidth]{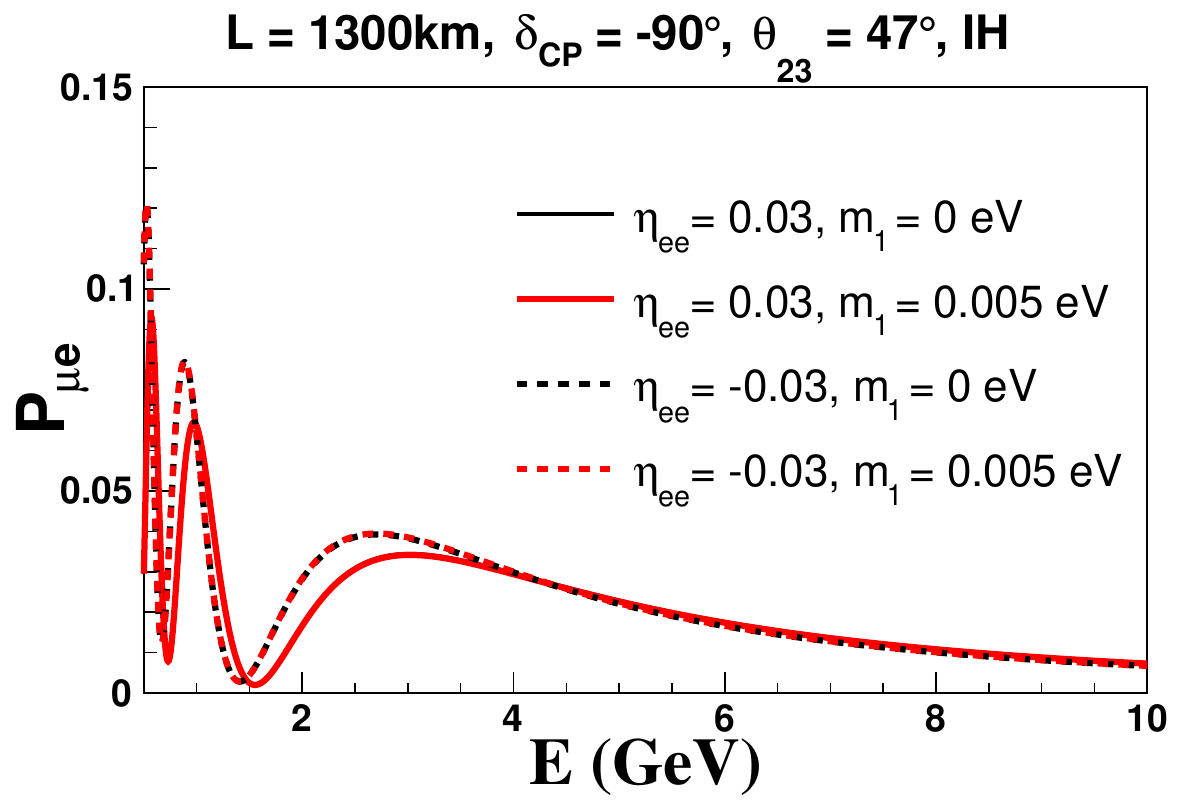}
\includegraphics[height=5cm,width=0.32\linewidth]{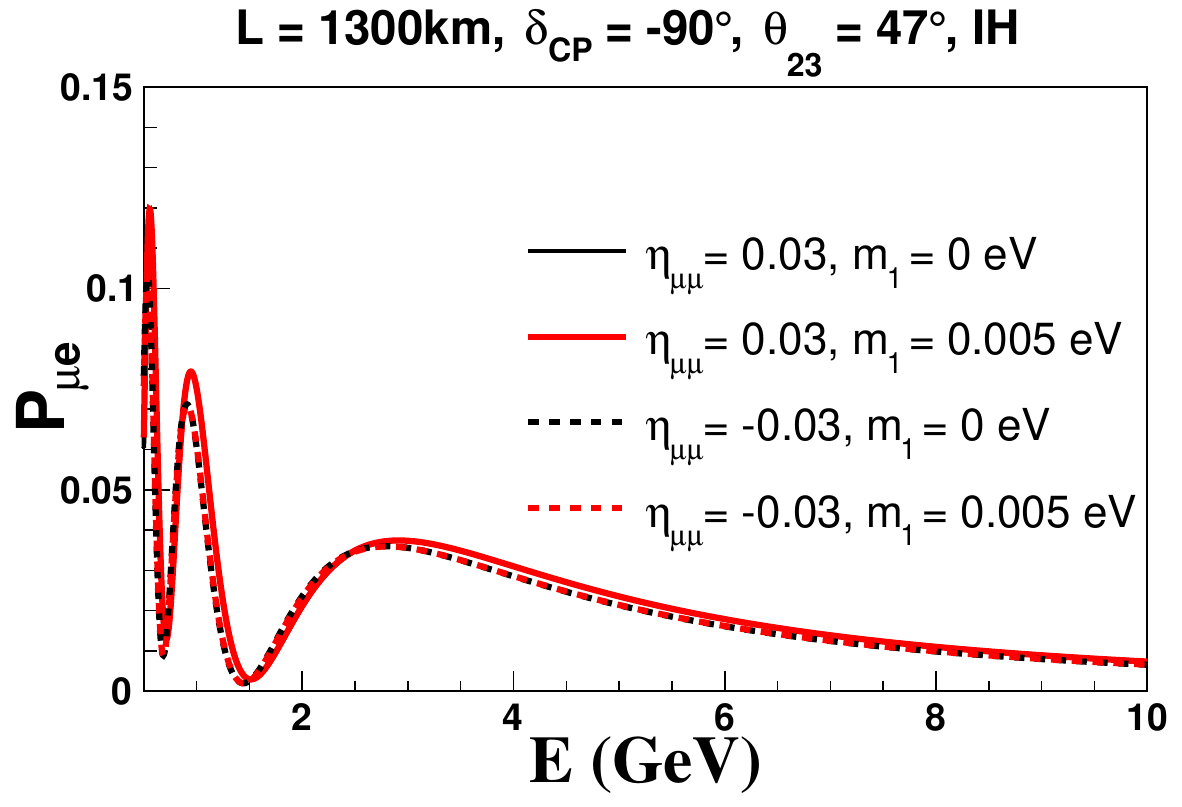}
\includegraphics[height=5cm,width=0.32\linewidth]{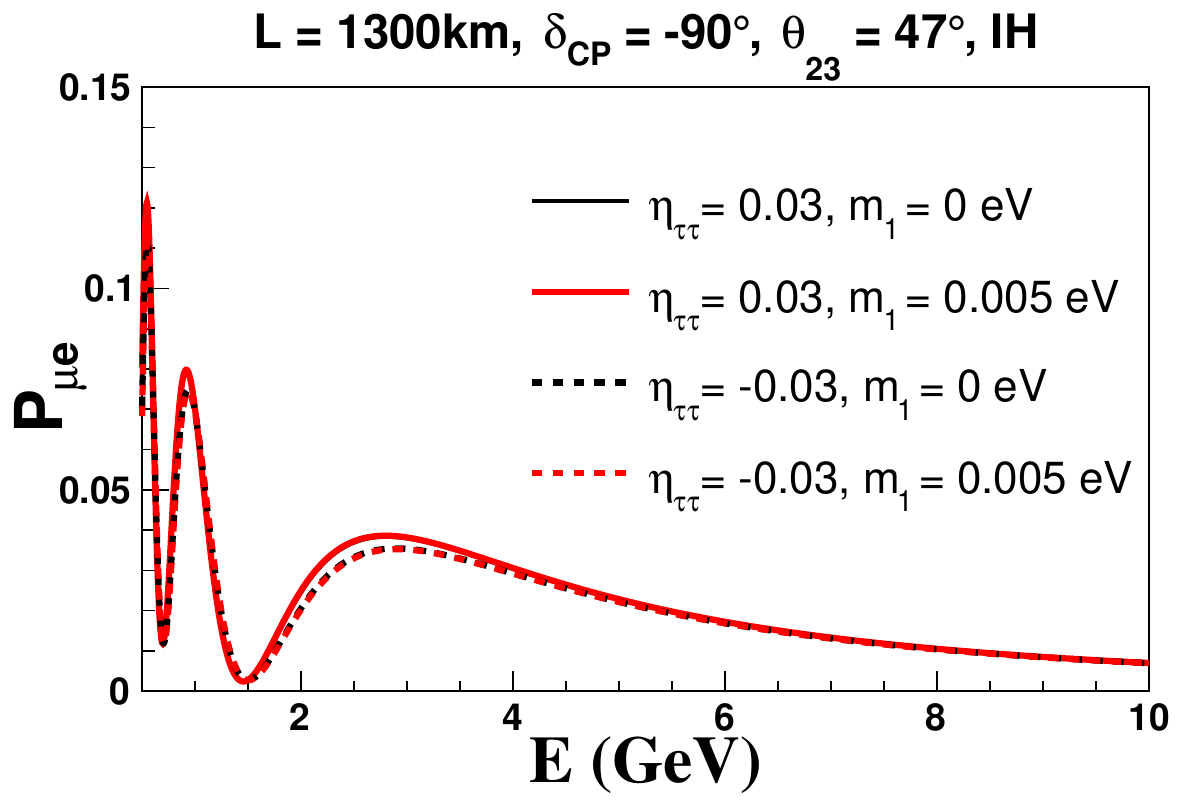}
\caption{Appearance probabilities for different choices of $m_\ell$ with $\eta_{ee}=\pm 0.03$ (left-panel), $\eta_{\mu\mu}= \pm 0.03$ (middle-panel) and $\eta_{\tau\tau}= \pm 0.03$ (right-panel) for normal hierarchy (top-panel) and inverted hierarchy (bottom-panel). The solid (dashed) line represents a positive (negative) value of $\eta_{\alpha\alpha}$. Also, we consider $\delta_{CP}$ = -$\pi$/2 and $\theta_{23} = 47^\circ$ to generate the plots.} 
\label{fig:app_prob_eta_ee_mm_tt}
\end{figure*}

\begin{multicols}{2}
\noindent We see a non-trivial dependence of the probability contributions on the absolute neutrino masses. We observe that the probability contributions can be expressed in factors of $(m_2-m_1)$, $(m_1+m_2)$, $(m_1 c_{12}^{2}+m_2 s_{12}^{2})$ and $m_3$. The contribution is significantly different for the three diagonal SNSI elements. In the following section, we explore the impact of the SNSI element on the numerically calculated oscillation probabilities.

\subsection{Impact on neutrino oscillation probabilities}\label{sec:Pmue} 
In Fig. \ref{fig:app_prob_eta_ee_mm_tt}, we explore the impact of different allowed values of $m_\ell$ on the numerically calculated appearance probability ($P_{\mu e}$) in the presence of SNSI elements $\eta_{ee}$ (left-panel), $\eta_{\mu\mu}$ (middle-panel) and $\eta_{\tau\tau}$ (right-panel) for DUNE. We have plotted the appearance probability $P_{\mu e}$ for varying neutrino energies in the range 0.5 - 10 GeV. The solid colored lines represent the positive value of SNSI parameters, whereas the dashed lines correspond to negative values. Additionally, the top panel corresponds to the true normal hierarchy and the bottom panel represents the true inverted hierarchy. The values of oscillation parameters used are as described in table \ref{tab:param_val}.

\begin{itemize}
    \item For NH (top-panel), we observe a marginal deviation as the scale of absolute neutrino mass shifts to a higher value. The effect on $P_{\mu e}$ for different choices of $m_{1}$ are different for both positive and negative $\eta_{\alpha\alpha}$. In presence of positive (negative) $\eta_{\mu\mu}$, we see a nominal suppression (enhancement) of probabilities for higher values of neutrino masses. We also note slight suppression (enhancement) for positive (negative) $\eta_{\tau\tau}$. 

    \item For IH (bottom-panel), no significant variation with the lightest neutrino mass is observed for all the $\eta_{\alpha\alpha}$ element. However, for positive (negative) values of $\eta_{ee}$ and $\eta_{\mu\mu}$, a slight shift in the energy of the oscillation peak is noted.
\end{itemize}
We note that the presence of SNSI parameters can substantially impact the neutrino oscillation probabilities. This provides a unique opportunity to probe SNSI effects, offering a potential avenue to place constraints on the neutrino mass.

\begin{figure*}[!b] 
\centering
\includegraphics[height=5cm,width=0.32\linewidth]{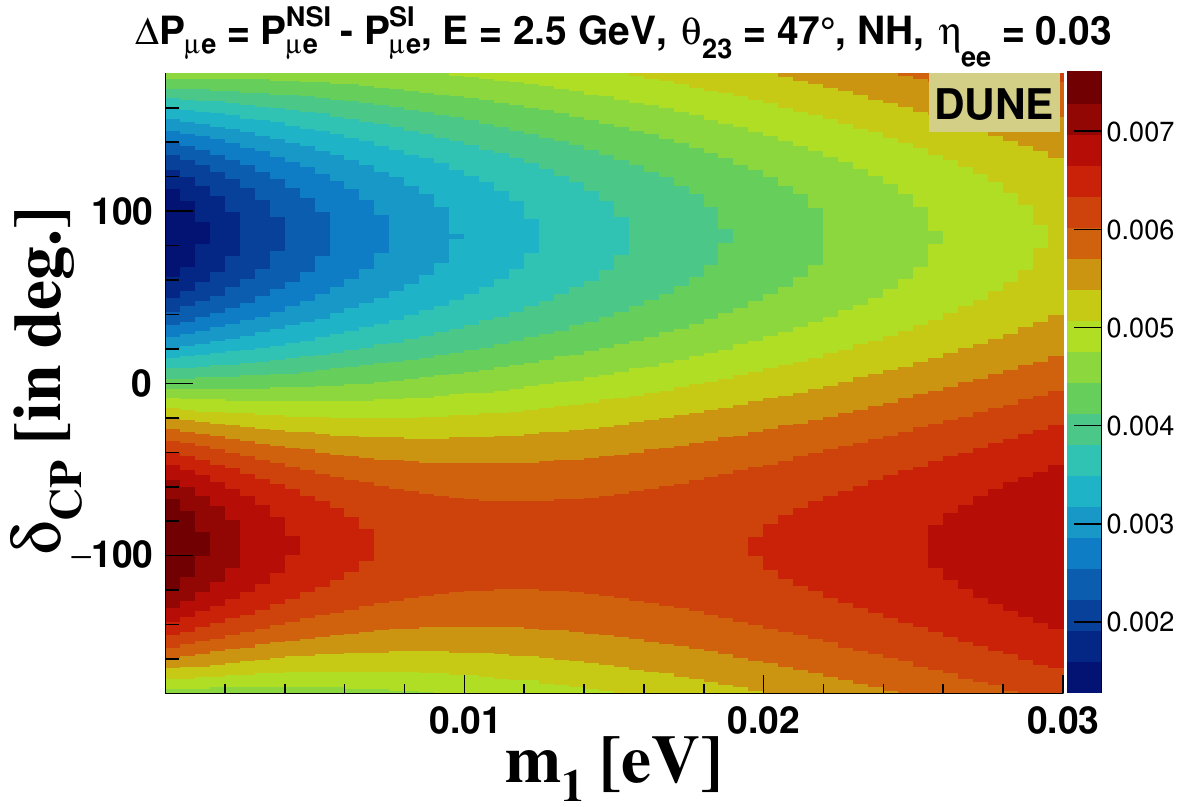}
\includegraphics[height=5cm,width=0.32\linewidth]{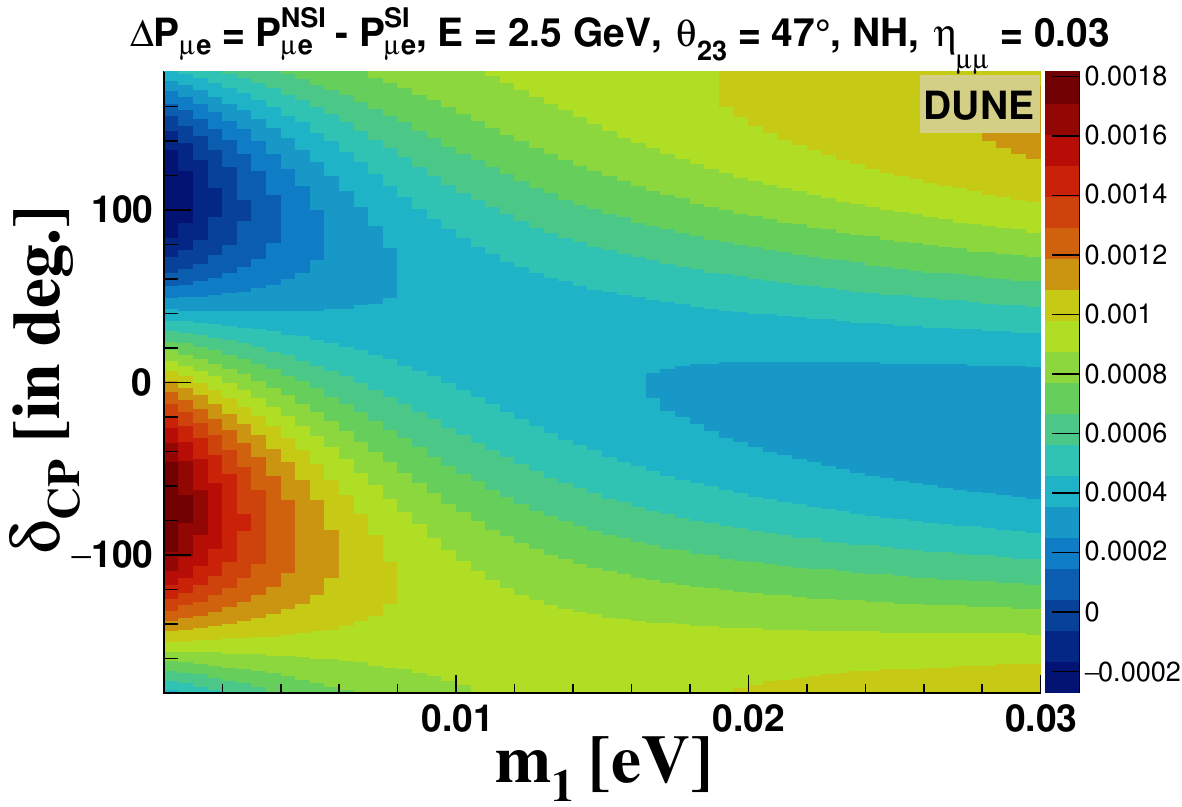}
\includegraphics[height=5cm,width=0.32\linewidth]{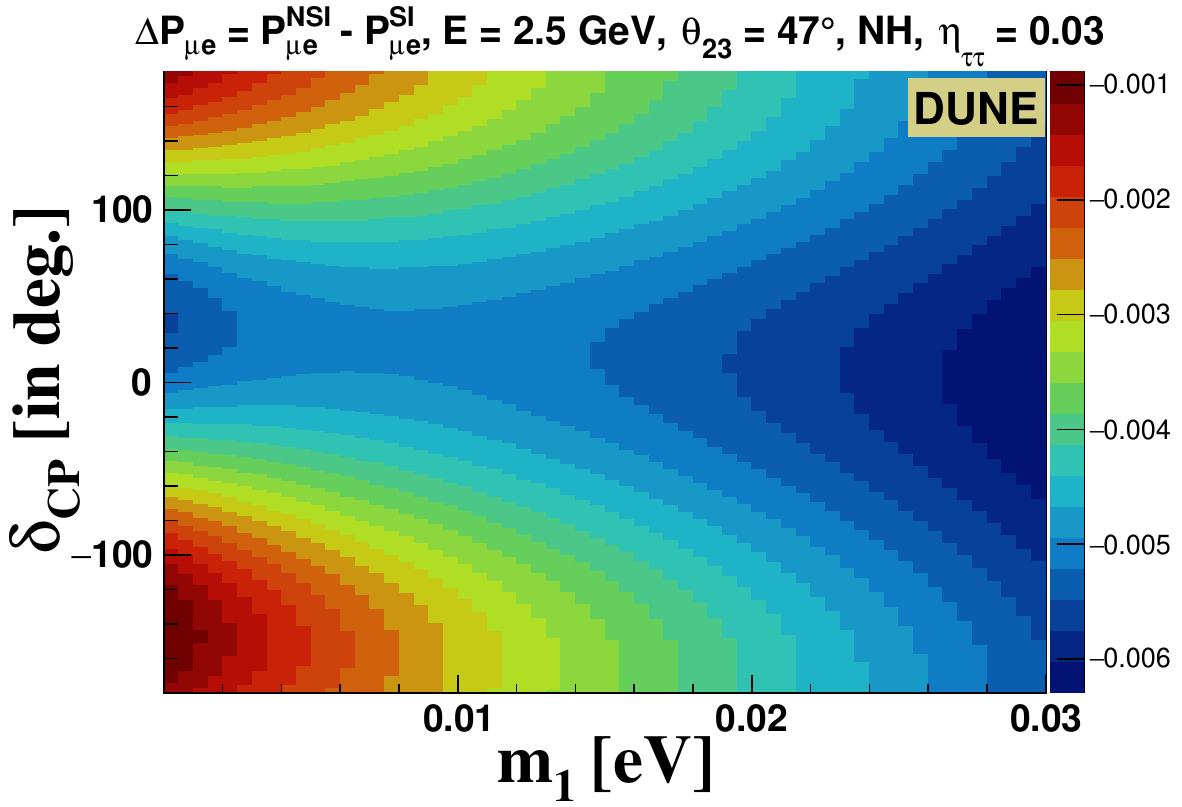}
\includegraphics[height=5cm,width=0.32\linewidth]{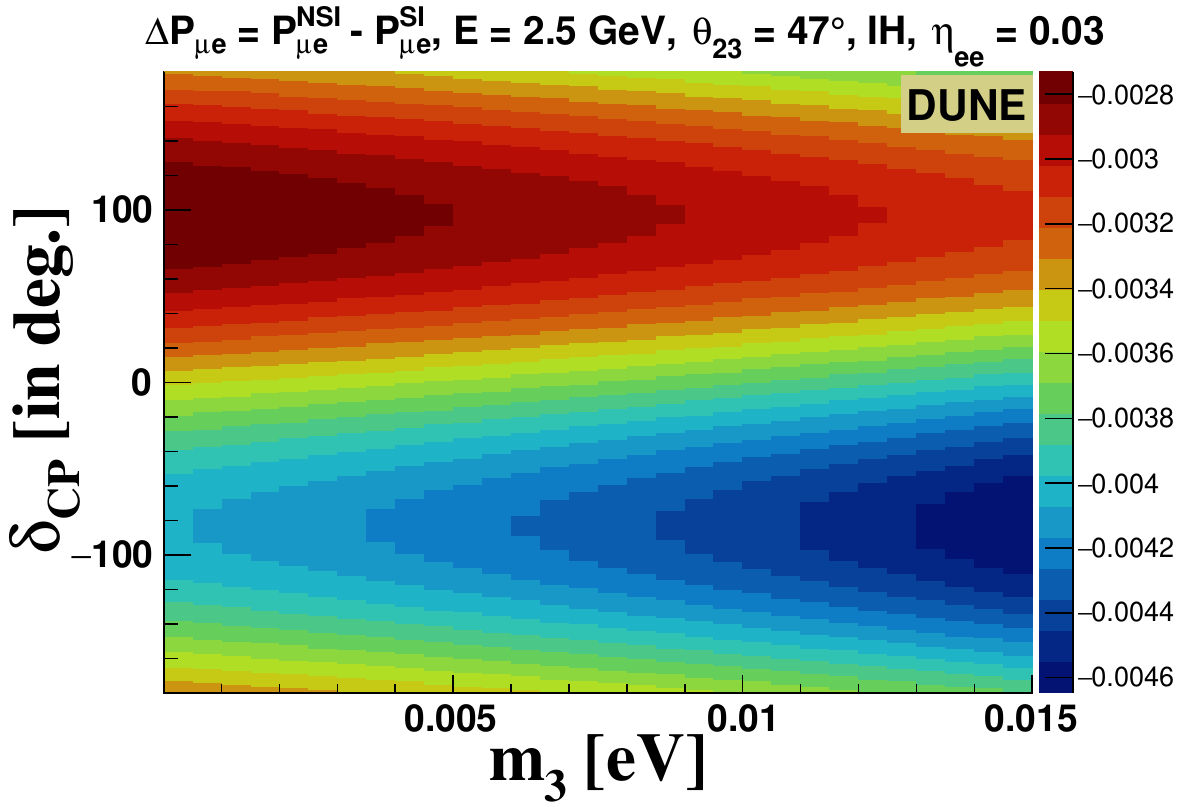}
\includegraphics[height=5cm,width=0.32\linewidth]{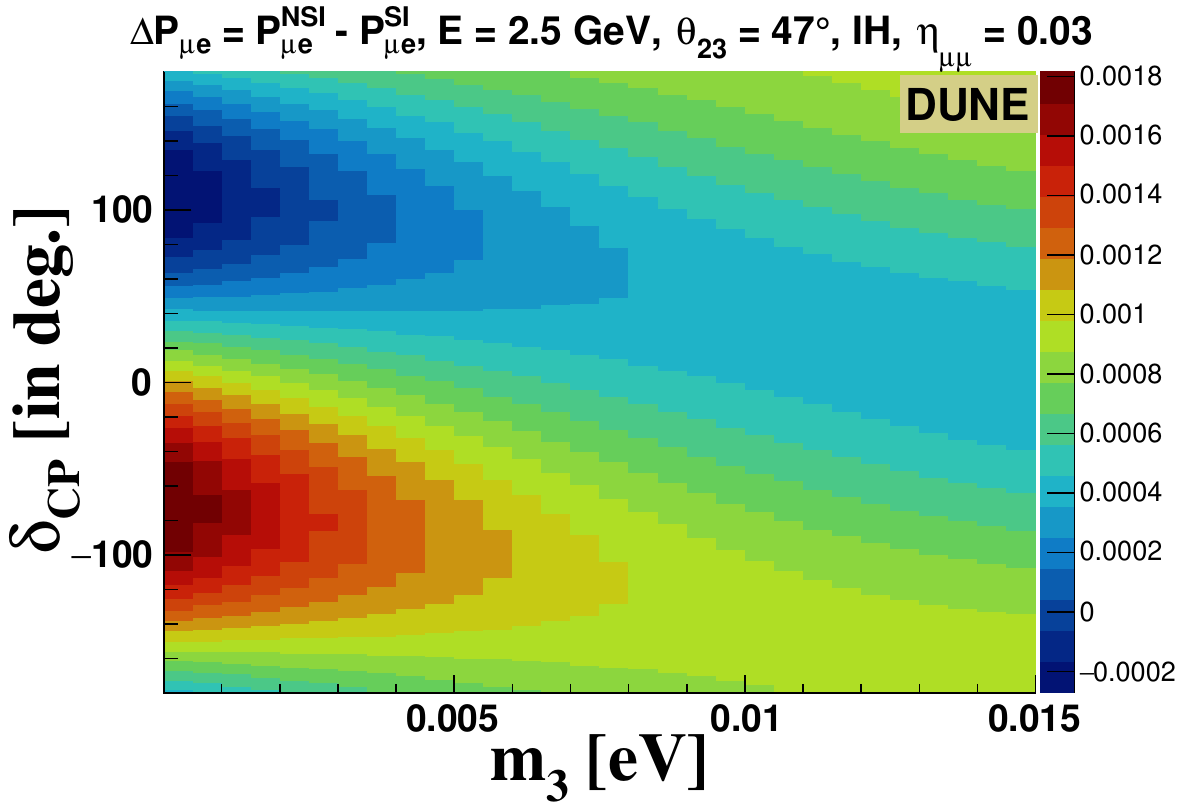}
\includegraphics[height=5cm,width=0.32\linewidth]{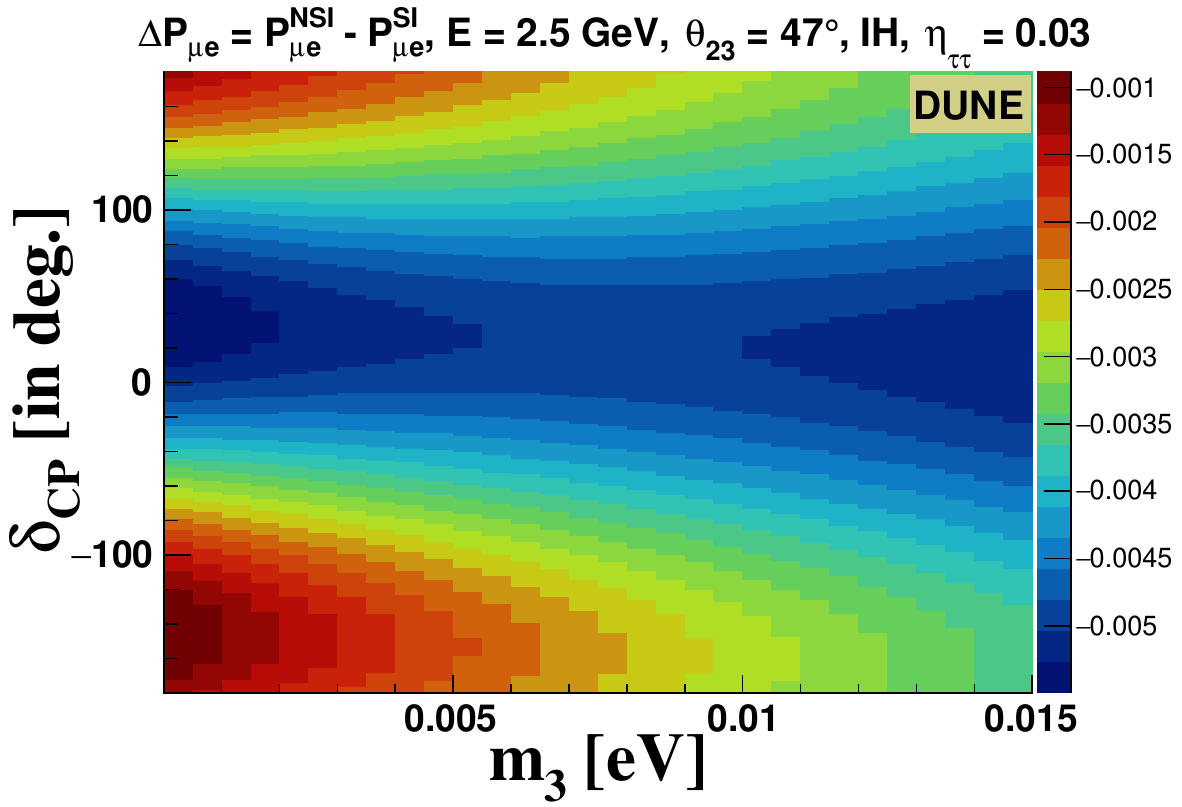}
\caption{Plot of $\Delta P_{\mu e} = P_{\mu e}^{NSI} - P_{\mu e}^{SI}$ in a two-dimensional histogram in $(\delta_{CP}$--$m_\ell)$ plane  with $\eta_{ee}= 0.03$ (left-panel), $\eta_{\mu\mu}= 0.03$ (middle-panel) and $\eta_{\tau\tau}= 0.03$ (right-panel) at DUNE. The top--panel (bottom--panel) represents  NH (IH) for neutrino energy = 2.5 GeV, $\theta_{23}$ = $47^\circ$. The left, middle and right panels correspond to SNSI elements $\eta_{ee}$, $\eta_{\mu\mu}$ and $\eta_{\tau\tau}$.}
\label{fig:dcp-vs-m}
\end{figure*}

\subsection{Exploring the $(\delta_{CP}$ - $m_\ell)$ parameter space}\label{sec:Pmue_space}
We then explore the variation of neutrino oscillation probability channel ($P_{\mu e}$) as a function of $\delta_{CP}$ and the lightest neutrino mass for both normal and inverted hierarchy. To quantify the impact of SNSI elements on the oscillation probabilities, we define a parameter $\Delta P_{\mu e}$ as,
\begin{equation}
\Delta P_{\mu e} = P_{\mu e}^{NSI} - P_{\mu e}^{SI}    
\end{equation}
where $P_{\mu e}^{NSI}$ and $P_{\mu e}^{SI}$ are the $\nu_{e}$-appearance probabilities with and without SNSI respectively. In Fig. \ref{fig:dcp-vs-m}, we have explored the effect of SNSI with varying $\delta_{CP}$ and $m_\ell$ within the allowed region. The values of the other oscillation parameters are listed in table \ref{tab:param_val}. We have also fixed the neutrino energy at $E=2.5$ GeV. We then vary the mass in the allowed range and $\delta_{CP}$ $\in$ $[-\pi,\pi]$. We have plotted the numerically calculated values of $\Delta P_{\mu e}$ for varying $\delta_{CP}$ and neutrino mass. We take the SNSI parameter $\eta_{ee}$, $\eta_{\mu\mu}$ and $\eta_{\tau\tau}$ in the left, middle and right panel respectively. We show the effects for true NH (IH) in the top (bottom) panel. We observe that,

\begin{itemize}
    \item For NH case, the element $\eta_{ee}$ enhances the probabilities for all values of $\delta_{CP}$. This enhancement is particularly pronounced in the negative $\delta_{CP}$
    region. The element $\eta_{\mu\mu}$ brings nominal changes as compared to $\eta_{ee}$ and $\eta_{\tau\tau}$. In contrast, the $\eta_{\tau\tau}$ element suppresses the probabilities for the complete $\delta_{CP}$ space [$-\pi$, $\pi$].
    \item In the Inverted Hierarchy (IH) scenario, both $\eta_{ee}$ and $\eta_{\tau\tau}$ suppress the probabilities over the full range of $\delta_{CP}$. The element $\eta_{\mu\mu}$ enhances the probabilities and the enhancement is higher for the negative half-plane of $\delta_{CP}$.
\end{itemize}

\begin{figure*}[!b]
  \centering
  \includegraphics[height=5cm,width=0.32\linewidth]{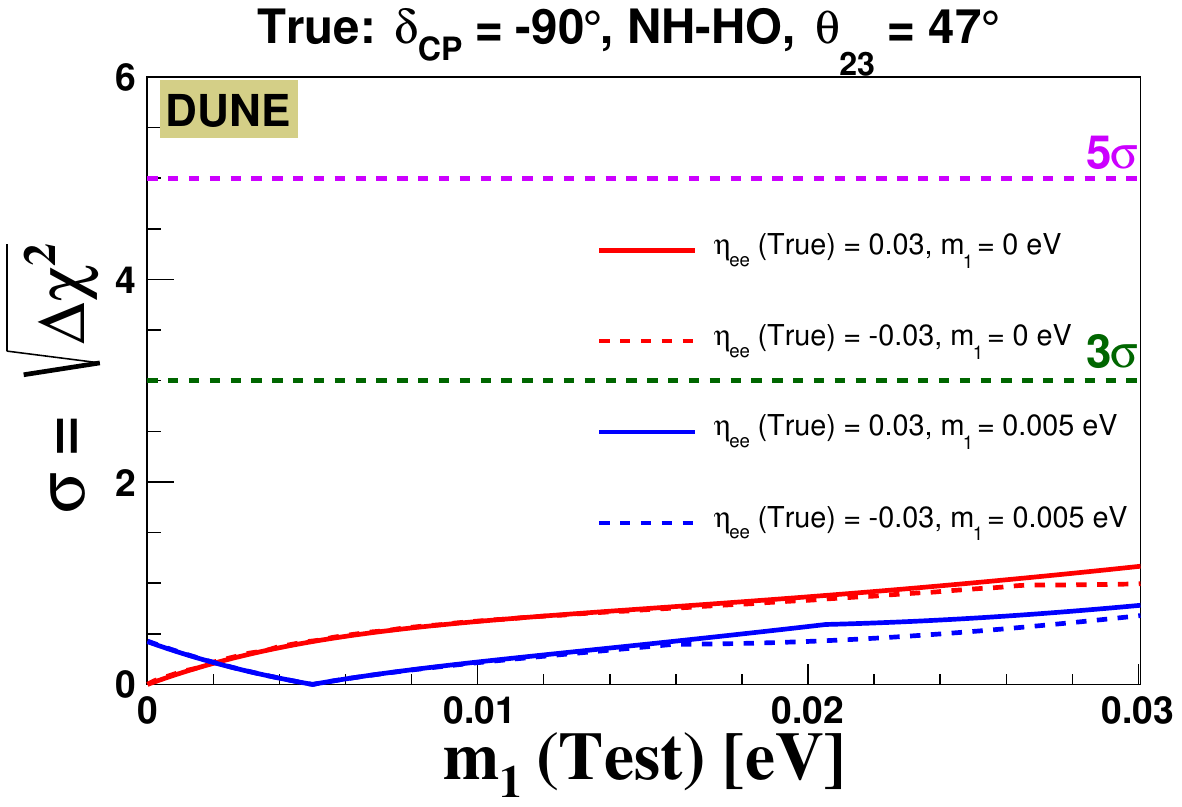}
  \includegraphics[height=5cm,width=0.32\linewidth]{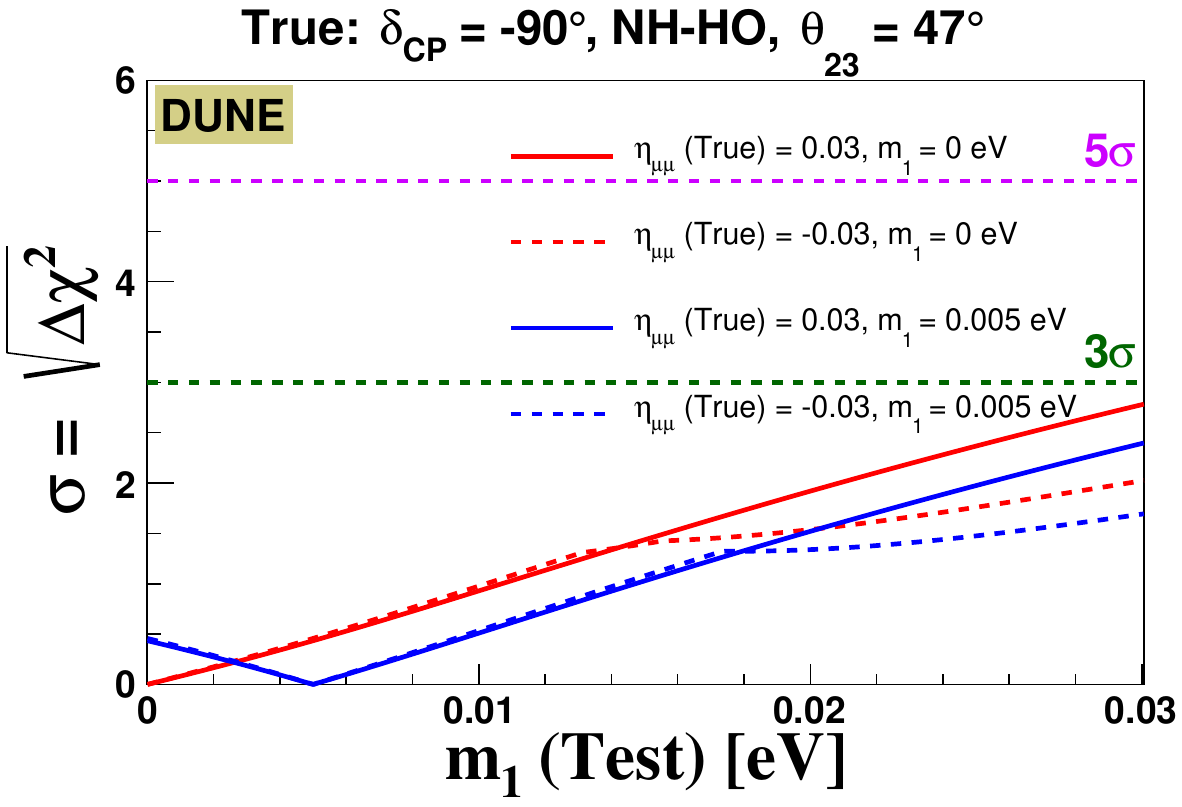}
  \includegraphics[height=5cm,width=0.32\linewidth]{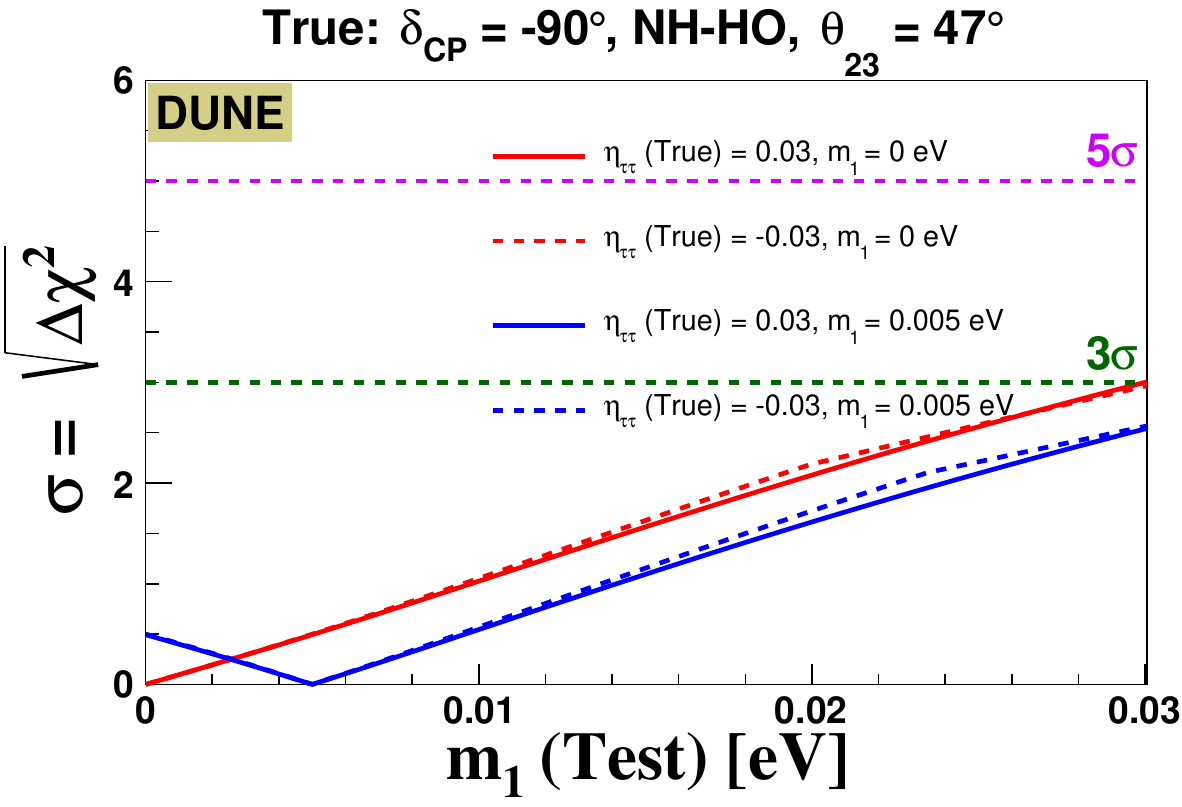}
 \includegraphics[height=5cm,width=0.32\linewidth]{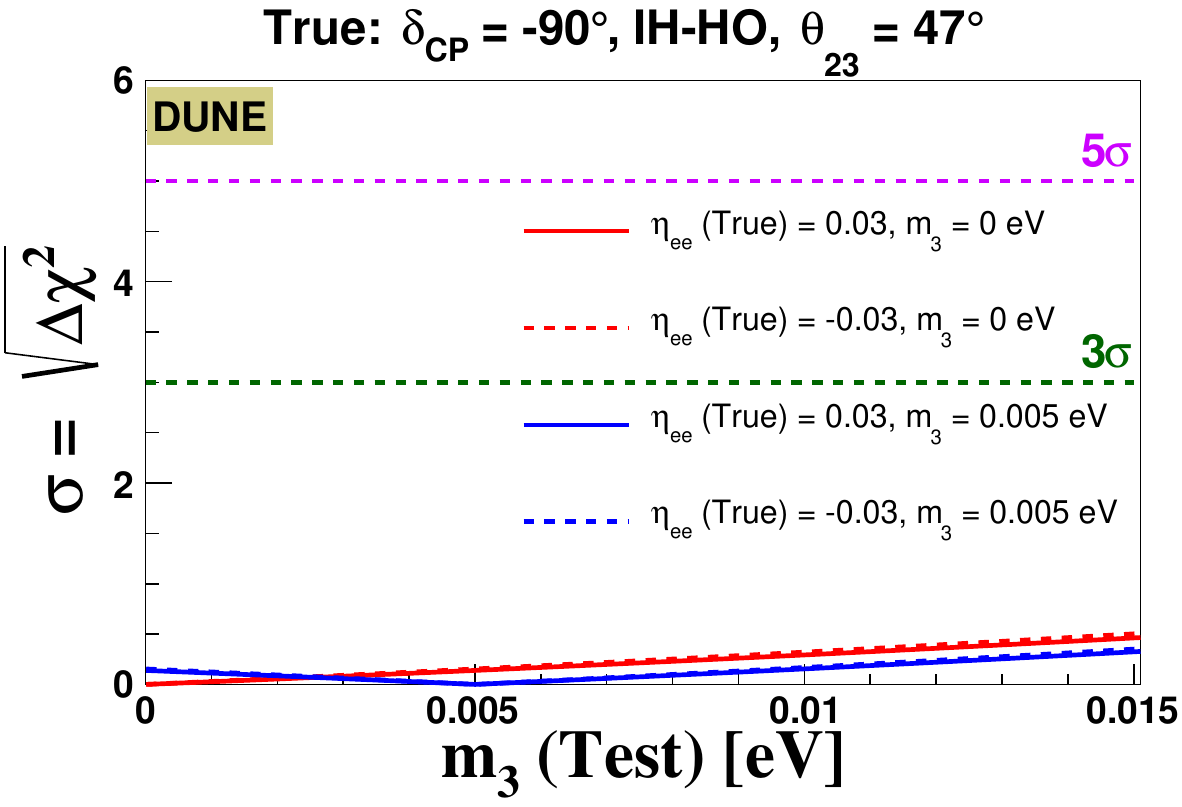}
  \includegraphics[height=5cm,width=0.32\linewidth]{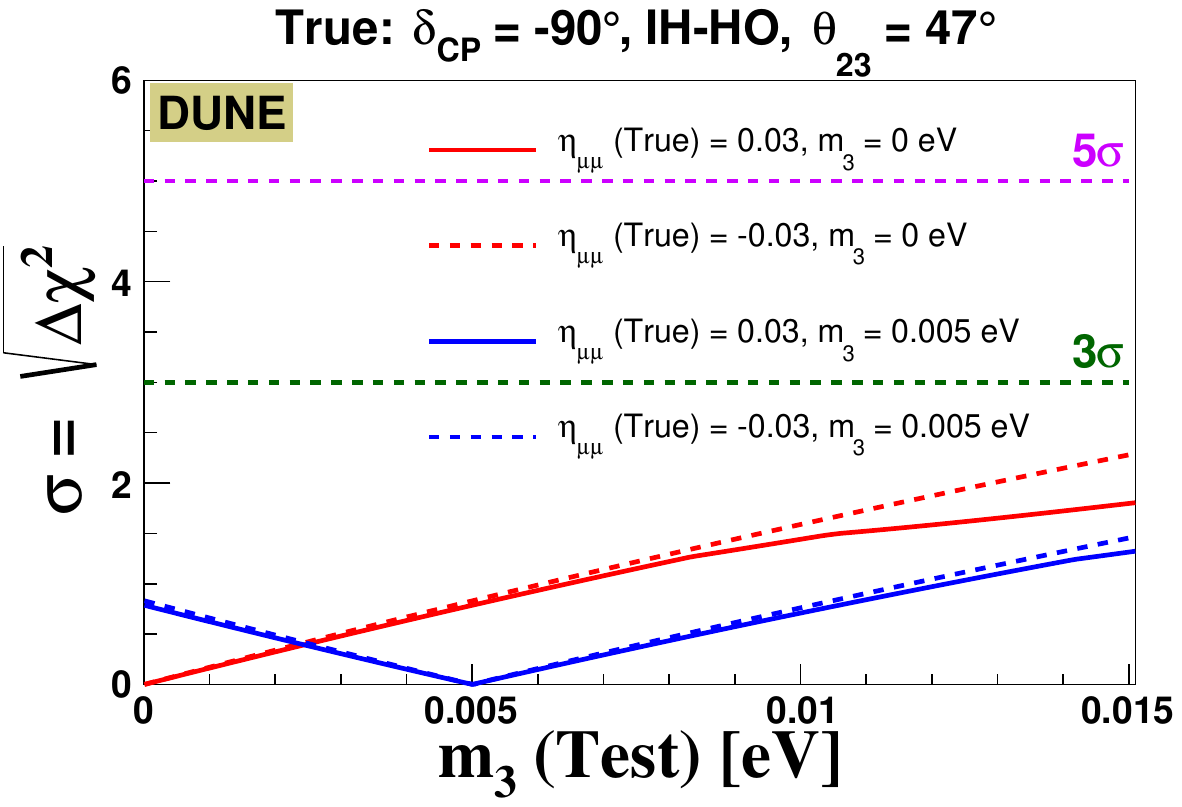}
  \includegraphics[height=5cm,width=0.32\linewidth]{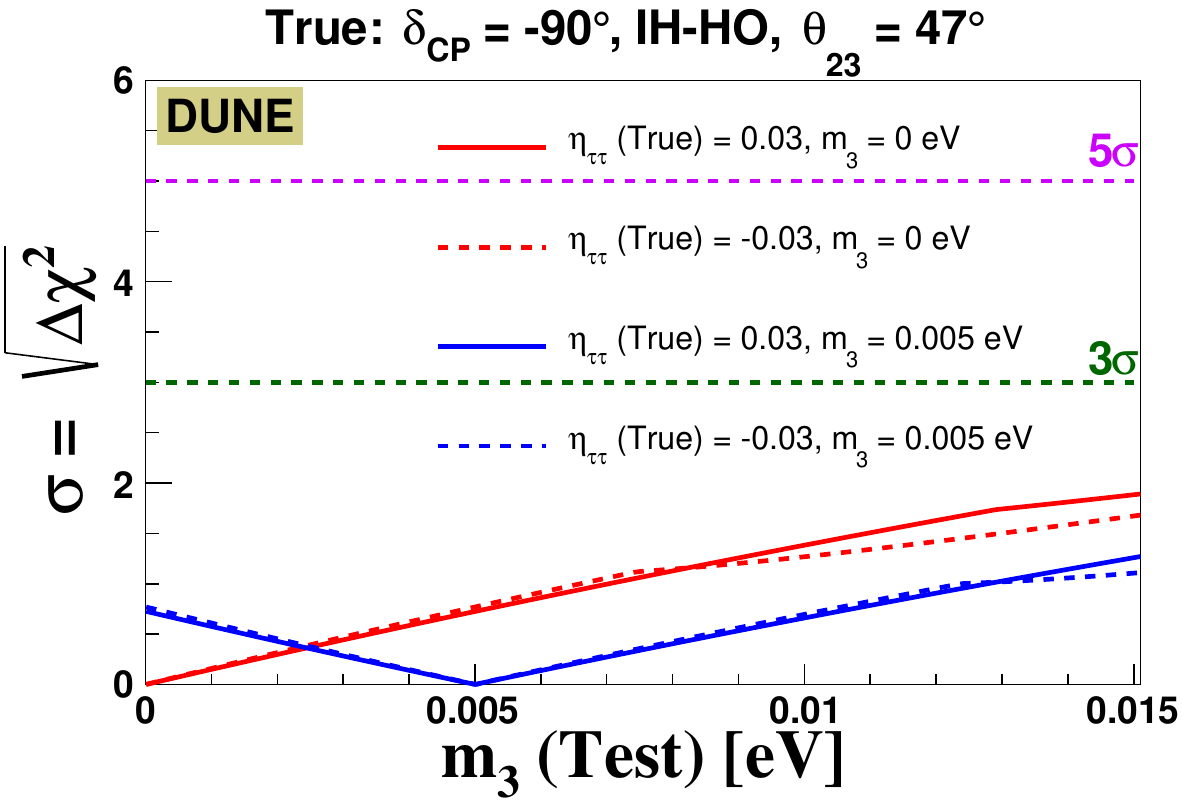}
\caption{Constrain on the lightest neutrino mass for normal (top-panel) and inverted (bottom-panel) hierarchy for the diagonal SNSI parameters $\eta_{ee}$ (left panel), $\eta_{\mu\mu}$ (middle-panel) and $\eta_{\tau\tau}$ (right-panel). The solid (dashed) lines represent positive (negative) parameter values. The red and blue color represents the case of the true lightest neutrino mass fixed at 0 eV and 0.005 eV for both hierarchies. The magenta and green dashed horizontal lines represent the $3\sigma$ and $5\sigma$ C.L.}
\label{fig:fix_chi2_2}
\end{figure*}

The exploration of the impact of SNSI on $P_{\mu e}$ motivates to explore the SNSI effects in constraining the neutrino mass. In the following sections, we discuss the constraints on the absolute neutrino mass in the presence of SNSI at DUNE.

\section{Constraining the lightest neutrino mass} \label{sec:bound_m}
We demonstrate the capability of DUNE towards constraining the lightest neutrino mass for both hierarchies. In Eq. \ref{eq:Hamil}, we see that the SNSI parameters contribute directly to the standard neutrino mass matrix which can be probed to place a bound on the neutrino masses. We define the sensitivity as $\Delta\chi^{2}$ which can help in constraining the masses.
\begin{align}\label{eq:delchi2}
\Delta\chi^{2} =min \Bigr[\chi^{2}&\left(\eta_{\alpha\alpha}^{test}\neq0,m_{\ell}^{test}\right)- \nonumber \\ 
&\chi^{2}\left(\eta_{\alpha\alpha}^{true}\neq0,m_{\ell}^{true}\neq0\right) \Bigr].
\end{align}
\noindent We investigate for two choices of lightest neutrino masses, i.e. $m_\ell$ is 0 eV and 0.005 eV as shown in Fig. \ref{fig:fix_chi2_2}. In the top panels, we have considered the case of true normal hierarchy whereas, in the bottom panels, we consider the true inverted hierarchy. We vary the test value of $m_\ell$ in the allowed range while fixing the true value of $m_\ell$ at 0 eV and 0.005 eV for both hierarchies. The values of the neutrino oscillation parameters are listed in table \ref{tab:param_val}. We have marginalized over the oscillation parameters $\theta_{23}$ and $\delta_{CP}$ in the fit data. The sensitivity $\sigma$ = $\sqrt{\Delta\chi^{2}}$ is plotted for an allowed range of the lightest neutrino mass, where the solid (dashed) lines represent a positive (negative) value of $\eta_{\alpha\alpha}$. We focus on the diagonal parameters $\eta_{ee}$ (left-panel), $\eta_{\mu\mu}$ (middle-panel) and $\eta_{\tau\tau}$ (right-panel) for NH (top-panel) and IH (bottom-panel) respectively. We have varied $m_{1}$ ($m_{3}$) for NH (IH) in the allowed region. In all the plots, the dashed magenta and green lines represent 5$\sigma$ and 3$\sigma$ CL. We list our observations as follows,

\begin{figure*}[!b]
  \centering
  \includegraphics[height=5cm,width=0.32\linewidth]{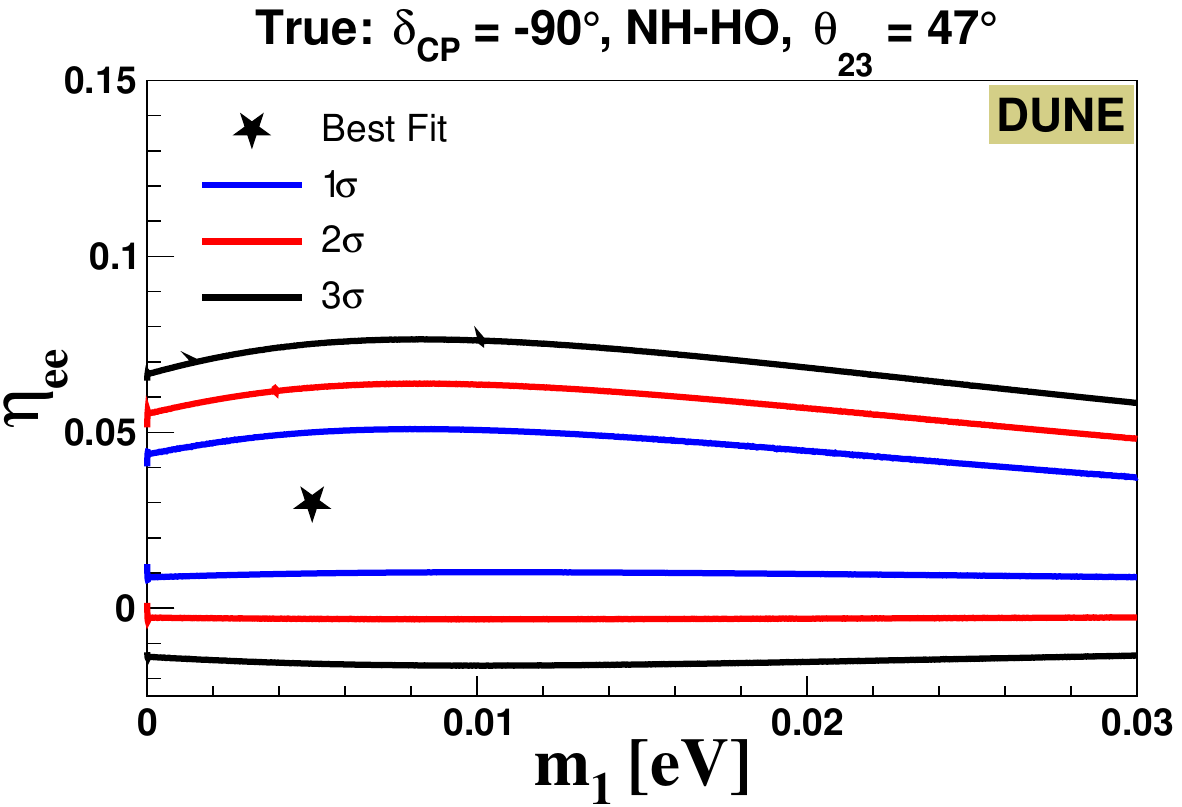}
  \includegraphics[height=5cm,width=0.32\linewidth]{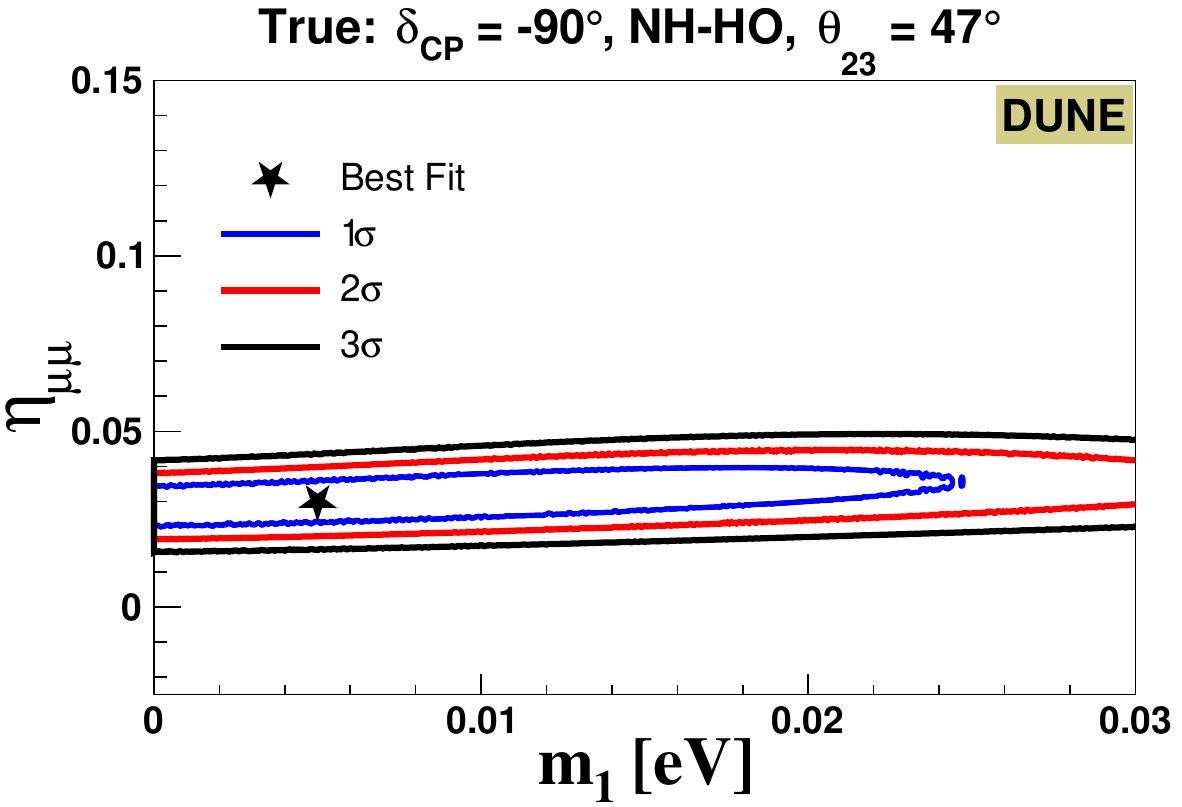}
  \includegraphics[height=5cm,width=0.32\linewidth]{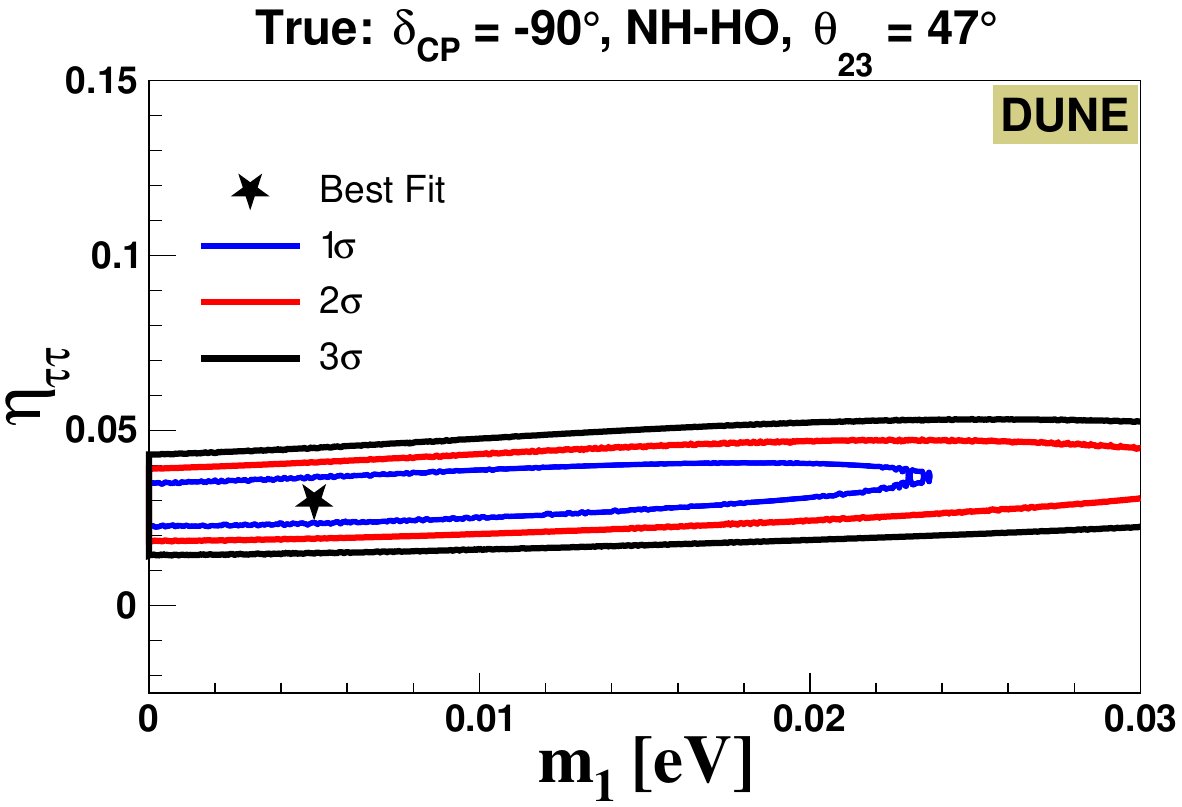}
 \includegraphics[height=5cm,width=0.32\linewidth]{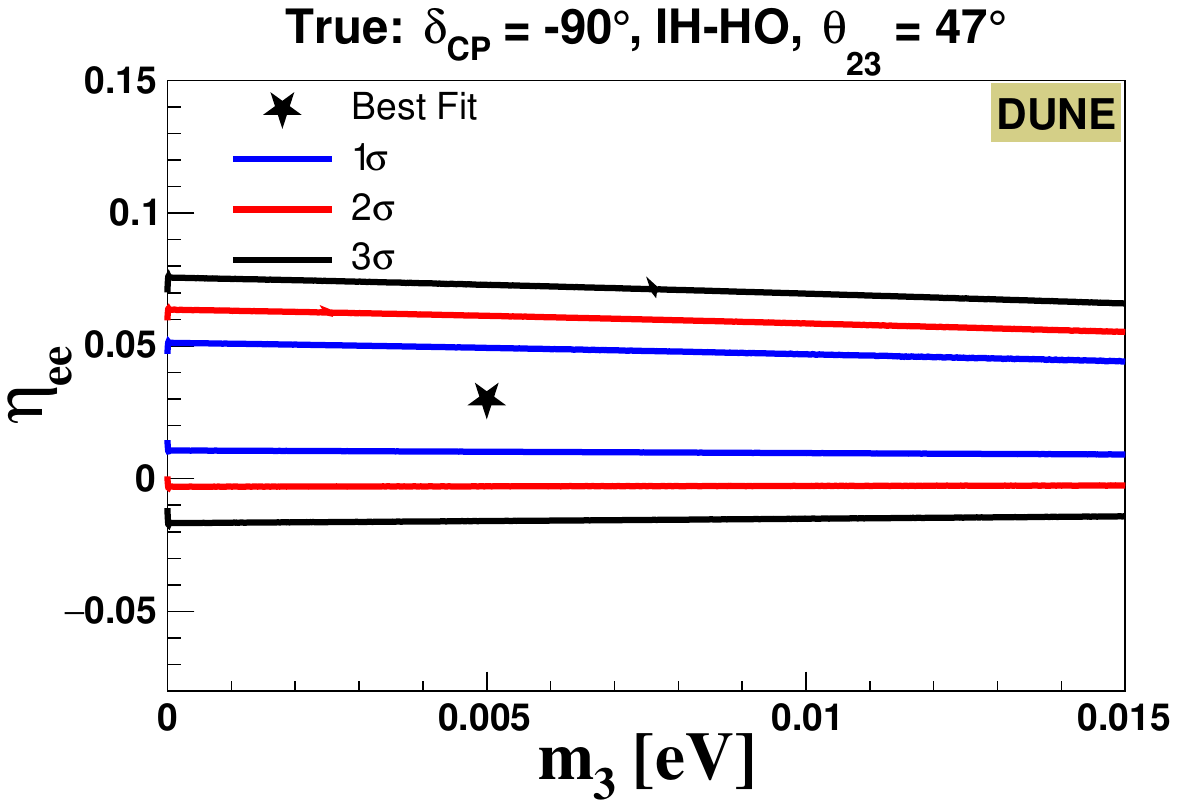}
  \includegraphics[height=5cm,width=0.32\linewidth]{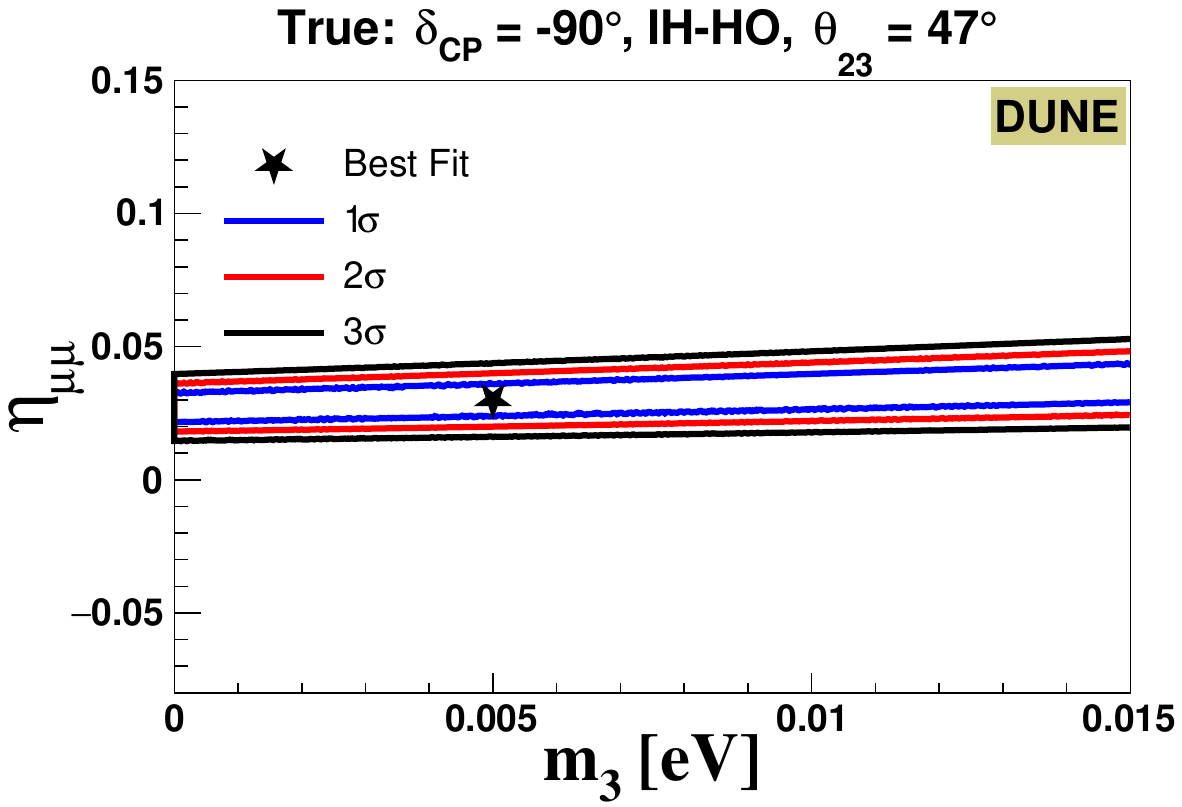}
  \includegraphics[height=5cm,width=0.32\linewidth]{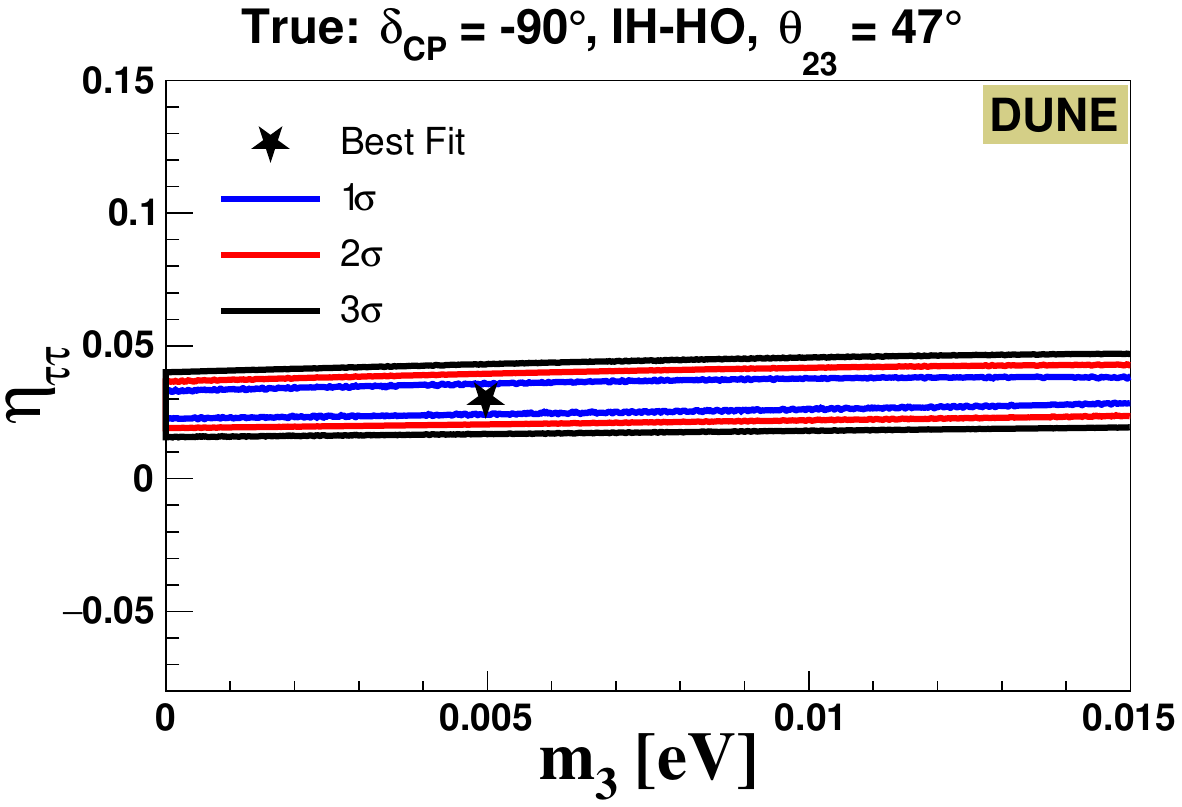}
\caption{\textbf{$\delta_{CP}=-90^{\circ}$, HO:} Allowed region of lightest neutrino mass for normal (top-panel) and inverted (bottom-panel) hierarchy for the diagonal SNSI parameters $\eta_{ee}$ (left-panel), $\eta_{\mu\mu}$ (middle-panel) and $\eta_{\tau\tau}$ (right-panel). The blue, red and magenta lines represent the 1$\sigma$, 2$\sigma$ and 3$\sigma$ confidence levels, respectively. The true value of $m_\ell$ is fixed at $0.005$ eV for NH and IH. The SNSI parameter $\eta_{\alpha\alpha}$ is fixed at 0.03. The best-fit point is represented by the solid black star.}
\label{fig:corr_chi2_1}
\end{figure*}

\begin{itemize}
    \item In Fig. \ref{fig:fix_chi2_2}, we present DUNE's capability to constrain the lightest neutrino mass under the assumption that the $m_\ell$ is fixed at 0 eV and 0.005 eV. This analysis aims to illustrate that even if the lightest neutrino mass is zero, DUNE remains sensitive in constraining the neutrino mass in the presence of certain SNSI parameters. The upper (lower) panel corresponds to NH (IH), while the left, middle, and right columns depict the cases for $\eta_{ee}$, $\eta_{\mu\mu}$, and $\eta_{\tau\tau}$ respectively.

    \item We note that the neutrino mass cannot be constrained at 3$\sigma$ CL for any chosen cases. However, we observe better constraining in presence of $\eta_{\mu\mu}$ and $\eta_{\tau\tau}$ compared to $\eta_{ee}$ for NH scenario (top-panel). A similar constraint on the mass is observed for $\eta_{\mu\mu}$ (middle panel) and $\eta_{\tau\tau}$ (right panel) elements. For all the cases shown, we observe that the constraining is slightly better for true NH than for IH. A positive (negative) $\eta_{ee}$ and $\eta_{\mu\mu}$ enhances (suppresses) the sensitivities. However, no significant change is observed with the change in sign of $\eta_{\tau\tau}$ for the NH case.
    
    \item For IH scenario (bottom-panel), the constraint on mass for $\eta_{\mu\mu}$ (middle panel) and $\eta_{\tau\tau}$ (right panel) is similar and significantly better compared to $\eta_{ee}$ (left panel). A negative (positive) $\eta_{\mu\mu}$ enhances (suppresses) the sensitivity. However, a positive $\eta_{\tau\tau}$ leads to a nominally better constraining compared to a negative $\eta_{\tau\tau}$.

    \item We note that the lightest neutrino mass is better constrained in the presence of $\eta_{\mu\mu}$ and $\eta_{\tau\tau}$ for both the hierarchies.
\end{itemize}

In the presence of SNSI, we observe that the neutrino mass can be significantly constrained for both neutrino mass hierarchies. In the following section, we will demonstrate the sensitivity to neutrino mass constraints in the SNSI versus neutrino mass plane. This analysis will provide a clearer understanding of how various diagonal SNSI parameters can constrain the neutrino mass.

\begin{figure*}[!b]
  \centering
  \includegraphics[height=5cm,width=0.32\linewidth]{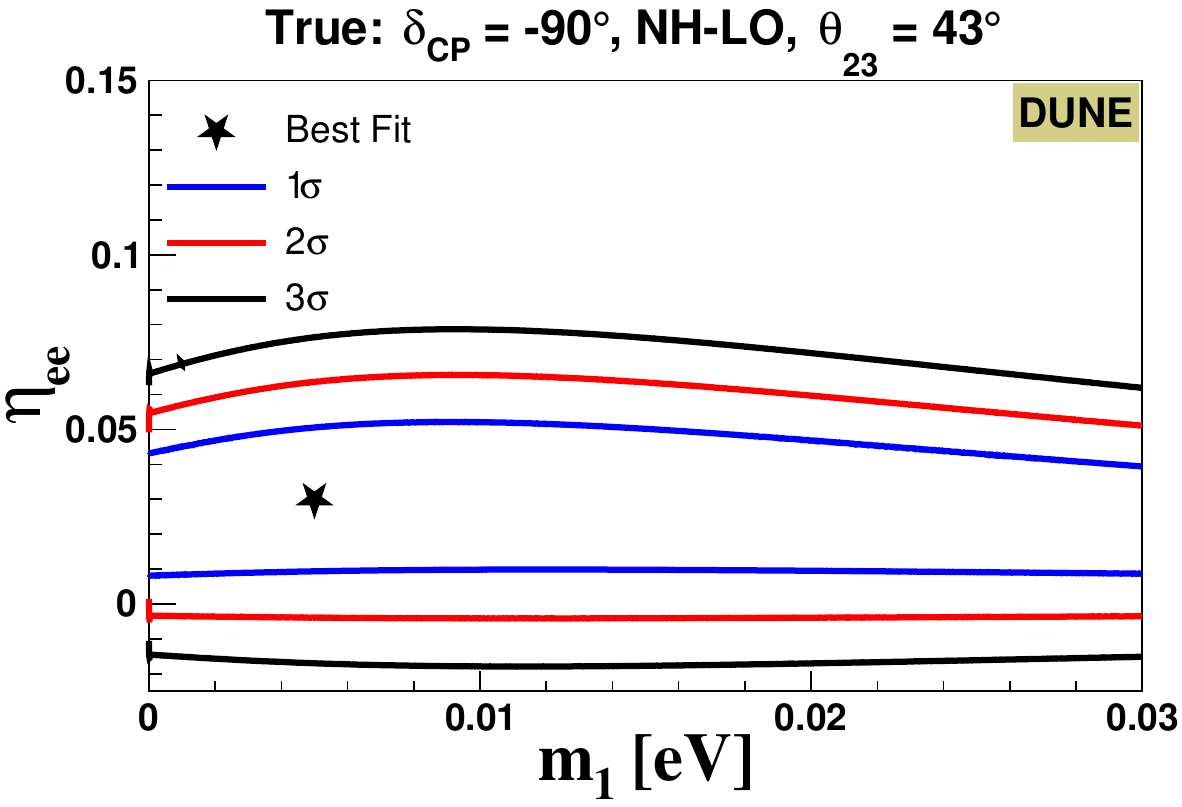}
  \includegraphics[height=5cm,width=0.32\linewidth]{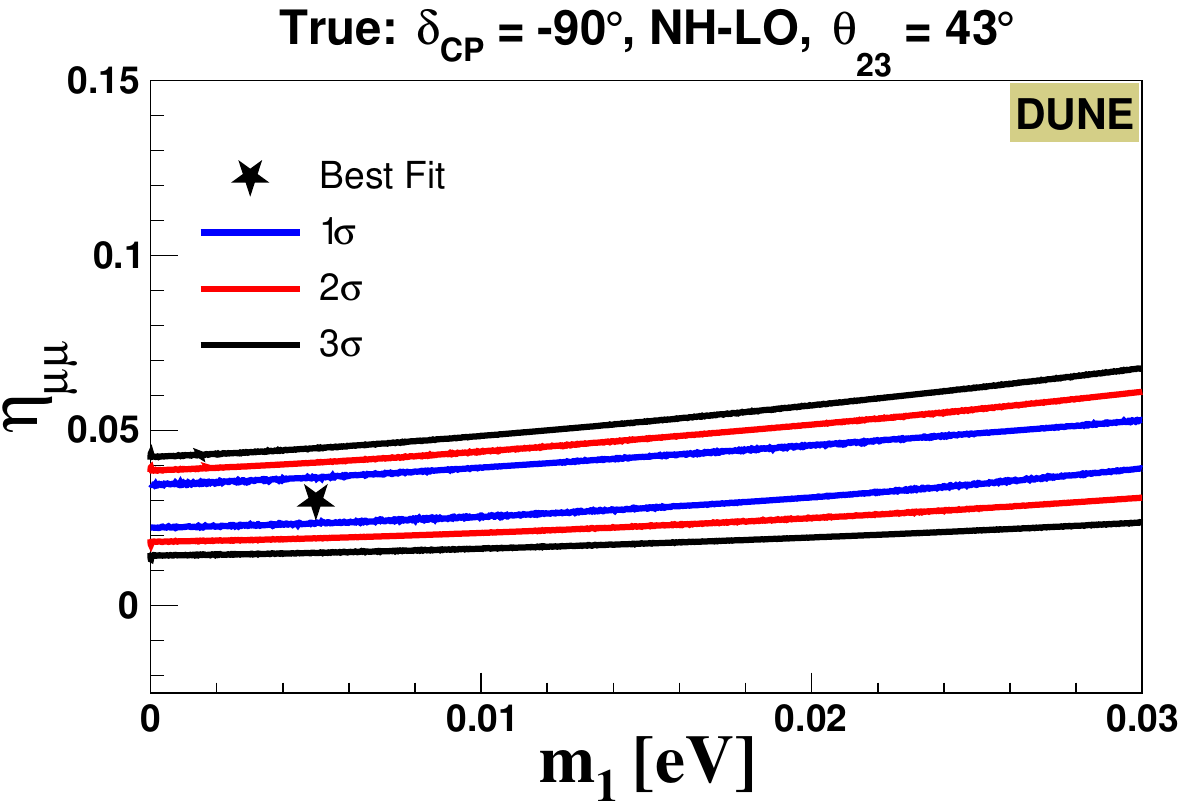}
  \includegraphics[height=5cm,width=0.32\linewidth]{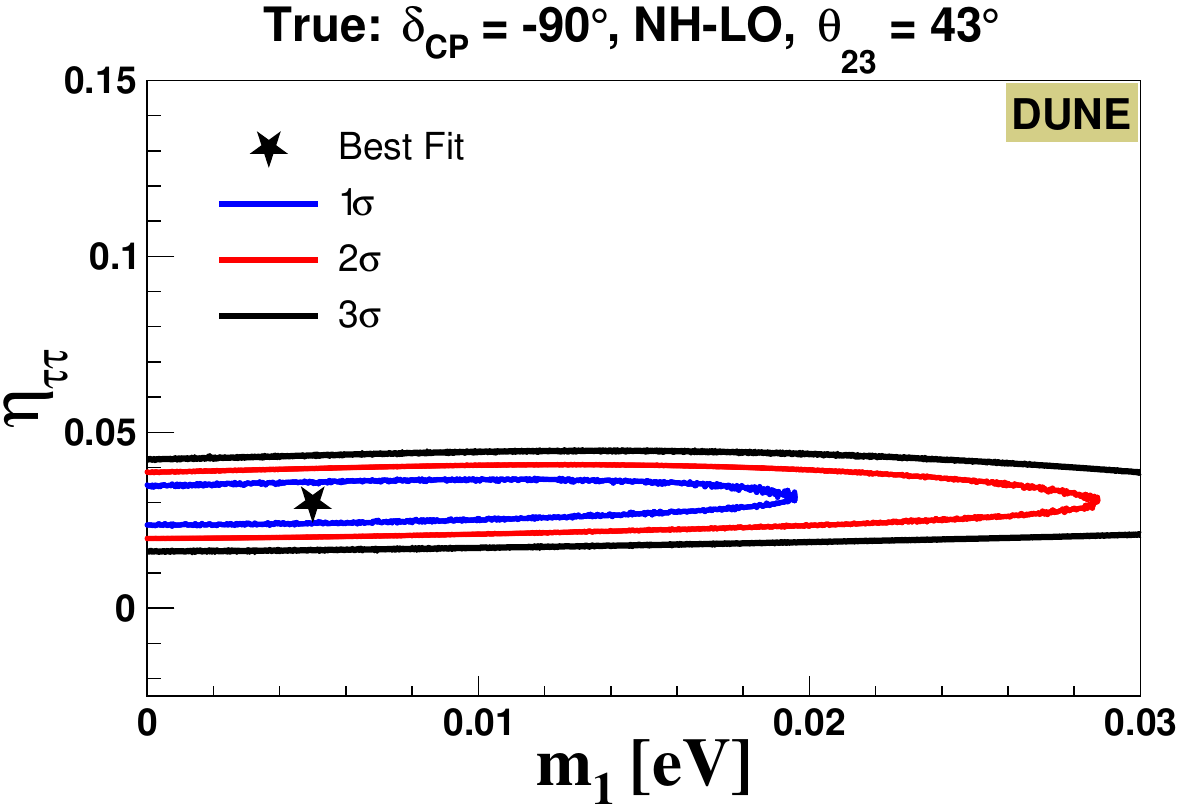}
 \includegraphics[height=5cm,width=0.32\linewidth]{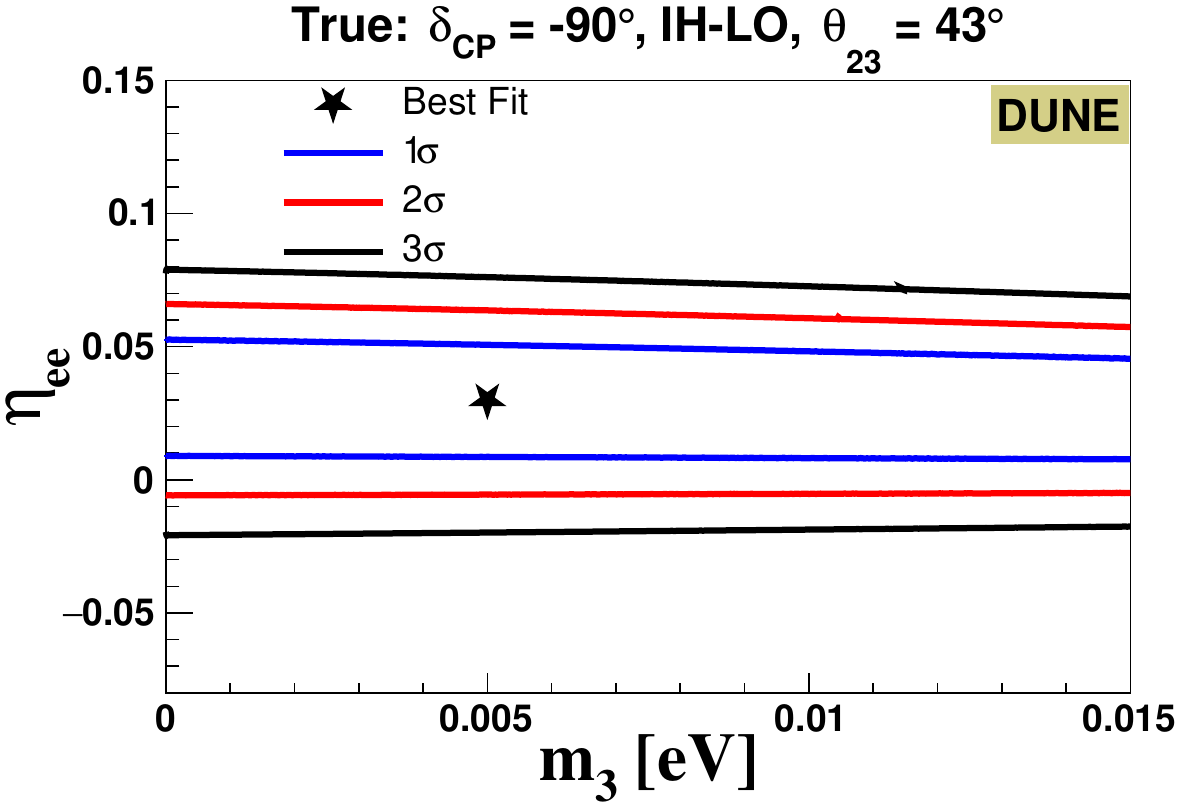}
  \includegraphics[height=5cm,width=0.32\linewidth]{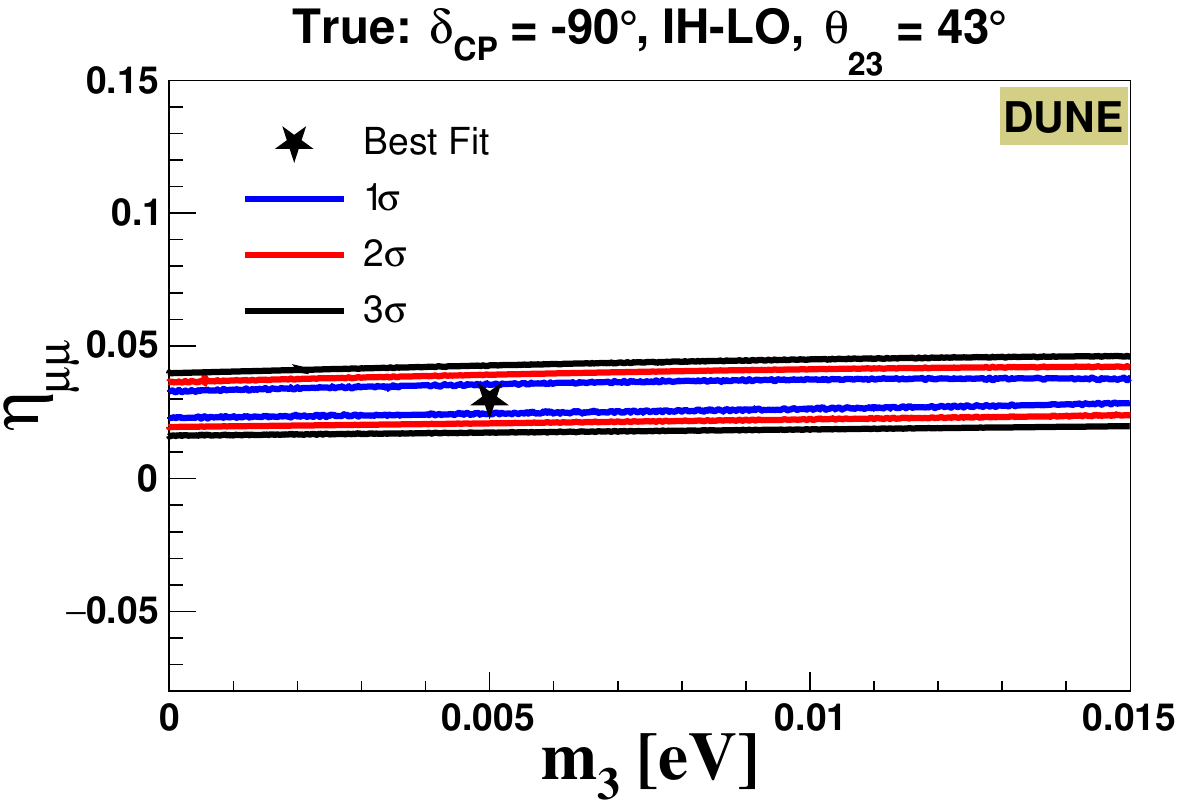}
  \includegraphics[height=5cm,width=0.32\linewidth]{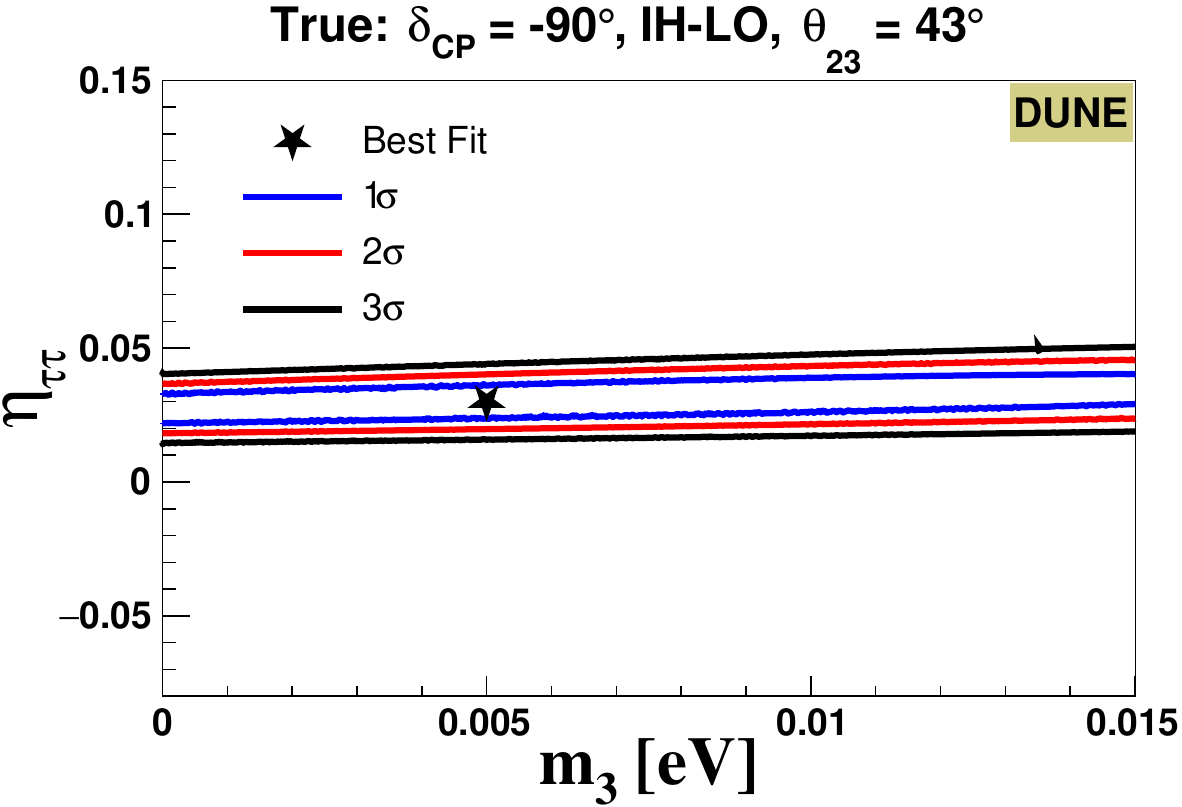}
\caption{\textbf{$\delta_{CP}=-90^{\circ}$, LO:} Allowed region of lightest neutrino mass for normal (top-panel) and inverted (bottom-panel) hierarchy for the diagonal SNSI parameters $\eta_{ee}$ (left-panel), $\eta_{\mu\mu}$ (middle-panel) and $\eta_{\tau\tau}$ (right-panel). The blue, red and magenta lines represent the 1$\sigma$, 2$\sigma$ and 3$\sigma$ confidence levels, respectively. The true value of $m_\ell$ is fixed at $0.005$ eV for NH and IH. The SNSI parameter $\eta_{\alpha\alpha}$ is fixed at 0.03. The best-fit point is represented by the solid black star.}
\label{fig:corr_chi2_2}
\end{figure*}

\section{Correlation in ($\eta_{\alpha\alpha}$ -- $m_\ell$) parameter space}\label{sec:correlation_m}

This study aims to place constraints on the lightest neutrino mass while considering the existence of SNSI in nature. In Fig. \ref{fig:corr_chi2_1} and Fig. \ref{fig:corr_chi2_2}, we show the allowed region for $1\sigma$, $2\sigma$ and $3\sigma$ CL in ($\eta_{\alpha\alpha}-m_\ell$) plane for DUNE  with $\delta_{CP}$ = $-90^\circ$ for higher (HO) and lower octant (LO) respectively. The remaining oscillation parameter values used in the study are listed in table \ref{tab:param_val}. The true values are represented by a solid black star. The test values of $\eta_{\alpha\alpha}$ are varied in [0, 0.15] and [-0.06, 0.15] for NH and IH respectively. The lightest neutrino mass is varied in the allowed range for NH and IH. We focus on the diagonal elements - $\eta_{ee}$ (left-panel), $\eta_{\mu\mu}$ (middle-panel) and $\eta_{\tau\tau}$ (right-panel) where the top (bottom) panel represents the NH (IH) case. In all the plots, the blue, red and black solid lines symbolize the $1\sigma$, $2\sigma$ and $3\sigma$ confidence regions respectively. The true point is represented by a black star. Our observations from the study are outlined as follows:

\begin{itemize}
    \item In Fig. \ref{fig:corr_chi2_1}, we observe that in the presence of $\eta_{ee}$, the lightest neutrino mass cannot be constrained for both mass hierarchies as observed in the left panel. The element $\eta_{\mu\mu}$ can constrain the lightest neutrino mass as $m_{1}$ $\in$ $[0.0, 0.025]$ eV at $1\sigma$ CL. For $\eta_{\tau\tau}$, we see a similar constrain on $m_{1}$ $\in$ $[0.0, 0.025]$ eV at $1\sigma$ CL. We also note that the mass cannot be constrained for any SNSI element in the inverted hierarchy (bottom panel). In all the cases we note that the constraint on $m_\ell$ worsens for the IH scenario. 

    \item In Fig. \ref{fig:corr_chi2_2}, the constraint on lightest neutrino mass for $\eta_{\mu\mu}$ (middle panel) weakens for NH. However, the constraining slightly improves in the presence of $\eta_{\tau\tau}$ element for the NH scenario. All the other cases show no significant changes. We observe that these results are mostly consistent regardless of $\theta_{23}$ octant except for $\eta_{\mu\mu}$.
    
    \item The findings of the study highlight the distinctive opportunity presented by SNSI to constrain the lightest neutrino masses. The investigation reveals that the exploration of SNSI offers a means to impose constraints on neutrino masses. We observe that the neutrino mass can be constrained only for true normal hierarchy. For higher octant, the constraining is possible for $\eta_{\mu\mu}$ and $\eta_{\tau\tau}$ whereas, for lower octant it is possible only for $\eta_{\tau\tau}$.
    
    \item Upon analyzing these results, it becomes evident that regardless of maximal and non-maximal CP violation, there exists a notable constraint on the neutrino mass when SNSI is incorporated into the model.
\end{itemize}

A similar pattern emerges when considering the scenario with $\delta_{CP}$ = $0^\circ$ for both $\theta_{23}$ octant configurations as shown in Appendix \ref{app:dcp0}. Our findings demonstrate minimal sensitivity to maximal or non-maximal $\delta_{CP}$ phase. This suggests a degree of robustness in our results. 

\section{Concluding Remarks } \label{sec:conclusion}
The upcoming neutrino experiments will be able to measure neutrino oscillation parameters with unparalleled precision. Nevertheless, the influence of subdominant contributions such as scalar non-standard interactions cannot be underestimated, as they possess the potential to significantly affect the detector sensitivities in these experiments. As SNSI directly modifies the neutrino mass term, it presents an interesting avenue to explore absolute neutrino masses within the neutrino oscillation sector. This study aims to demonstrate the use of neutrino oscillation data, in the context of SNSI, to constrain the lightest neutrino mass. This analysis is performed under the assumption that SNSI already exists in nature. 

We have constrained the lightest neutrino mass while considering the cosmological constraints imposed on the sum of neutrino masses. Our analysis reveals that the SNSI parameter can impose a constraint on the absolute mass of neutrinos. Specifically, the presence of $\eta_{\mu\mu}$ and $\eta_{\tau\tau}$ can put constraints on the lightest neutrino mass for normal mass hierarchy at 1$\sigma$ CL while considering higher octant of $\theta_{23}$. However, for lower octant only $\eta_{\tau\tau}$ is capable of putting bounds on the neutrino mass. The exploration of constraining neutrino mass holds significant importance, as it can provide crucial insights into the underlying mechanisms of neutrino mass generation.

\vspace{1cm}
\backmatter
\bmhead{Acknowledgements}
AM acknowledges the support of the project titled ``Indian Institutions - Fermilab Collaboration in Neutrino Physics" of IIT Guwahati funded by DST, Government of India. AS acknowledges the CSIR SRF fellowship (09/0796(12409)/2021-EMR-I) received from CSIR-HRDG. MMD acknowledges the Science and Engineering Research Board (SERB), DST for the grant CRG/2021/002961. The authors thank Debajyoti Dutta for his help and suggestions with the GLoBES framework. We also thank Dibya S. Chattopadhyay for the fruitful discussion and suggestions on the manuscript draft. We also thank the XXV DAE-BRNS HEP Symposium, held at IISER Mohali, the Workshop in High Energy Physics Phenomenology (WHEPP-XVII), held at IIT Gandhinagar and the NEUTRINO2024 conference held at Milan, Italy where the work has been discussed and presented.

\clearpage
\begin{table*}
    \centering
    \begin{tabular}{|c|c|c|c|}
    \hline
     \multicolumn{4}{|c|}{\rule{0pt}{15pt} Normal Hierarchy}\tabularnewline
    \hline 
    \rule{0pt}{12pt} Mass Squared Splitting & Best Fit value & Lower $(3\sigma)$ & Higher $(3\sigma)$\tabularnewline
    \hline 
    \hline 
    \rule{0pt}{12pt} $\Delta m_{21}^{2}(10^{-5}eV^{2})$ & 7.42 & 6.82 & 8.04\tabularnewline
    \hline 
    \rule{0pt}{12pt} $\Delta m_{31}^{2}(10^{-3}eV^{2})$ & 2.514 & 2.431 & 2.599\tabularnewline
    \hline 
    \multicolumn{4}{|c|}{\rule{0pt}{12pt} Inverted Hierarchy}\tabularnewline
    \hline 
    \rule{0pt}{12pt} Mass Squared Splitting & Best Fit value & Lower $(3\sigma)$  & Higher $(3\sigma)$\tabularnewline
    \hline 
    \hline 
    \rule{0pt}{12pt} $\Delta m_{21}^{2}(10^{-5}eV^{2})$ & 7.42 & 6.82 & 8.04\tabularnewline
    \hline 
    \rule{0pt}{12pt} $\Delta m_{32}^{2}(10^{-3}eV^{2})$ & -2.497 & -2.525 & 2.469\tabularnewline
    \hline 
    \end{tabular}
    \vspace{0.2cm}
    \caption{\small Mass-squared splitting for normal and inverted hierarchy with $3\sigma$ ranges.}
    \label{tab:dm_splitting_NO_IO}
\end{table*}

\begin{appendices}
\section{} \label{app:mass_range} 
Considering the best-fit values of mass-squared splittings from table \ref{tab:dm_splitting_NO_IO}, we obtain the upper limit of the lightest neutrino mass by taking the cosmological bound on the sum of neutrino masses. In table \ref{tab:m1_limit_NO}, we show the limit on the lightest neutrino masses $m_{1}$ and $m_{3}$ for NH and IH respectively, while following the cosmological bound $\sum m_{i}<0.12$ eV \cite{Planck:2018vyg,Dvorkin:2019jgs,FrancoAbellan:2021hdb,Wong:2011ip,Lesgourgues:2006nd}. We observe that $m_{1}$ and $m_{3}$ should be less than $0.03$ eV and $0.015$ eV for NH and IH respectively. The values of mass-squared splitting $\Delta m_{2k}^2$ and $\Delta m_{3k}^2$ for normal $(k=1)$ and inverted hierarchy $(k=2)$ with $3\sigma$ range are shown in table \ref{tab:dm_splitting_NO_IO}.

\begin{table*}
    \centering
    \renewcommand{\arraystretch}{1.5}
    \begin{tabular}{|c|c|c|c|c|c|c|c|}
    \hline
    \multicolumn{4}{|c|}{Normal Hierarchy} & \multicolumn{4}{c|}{Inverted Hierarchy}\tabularnewline
    \hline 
    Parameters & \multicolumn{3}{c|}{True Values} & \multicolumn{1}{c|}{Parameters} & \multicolumn{3}{c|}{True Values}\tabularnewline
    \hline 
    \hline 
    $m_{1}(eV)$ & 0.01 & 0.02828 & 0.03 & $m_{3}(eV)$ & 0.01 & 0.015 & 0.02\tabularnewline
    \hline 
    $m_{2}(eV)$ & 0.01319 & 0.02956 & 0.03121 & $m_{1}(eV)$ & 0.050227 & 0.051456 & 0.053130\tabularnewline
    \hline 
    $m_{3}(eV)$ & 0.0511 & 0.05757 & 0.05843 & $m_{2}(eV)$ & 0.050960 & 0.052172 & 0.053823\tabularnewline
    \hline 
    $\sum m_{i}(eV)$ & 0.0743 & 0.11541 & 0.11965 & $\sum m_{i}(eV)$ & 0.111187 & 0.118628 & 0.126953\tabularnewline
    \hline 
    \end{tabular}
        \vspace{0.2cm}
    \caption{\small Table representing the restricted range of neutrino masses for normal and inverted hierarchy.}
    \label{tab:m1_limit_NO}
\end{table*}

\begin{figure*}
    \centering
    \includegraphics[width=0.4\textwidth]{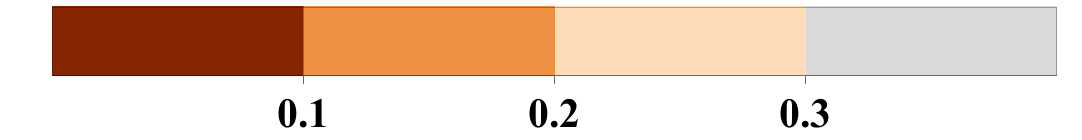}\\
    \includegraphics[width=0.32\textwidth, height=4.cm]{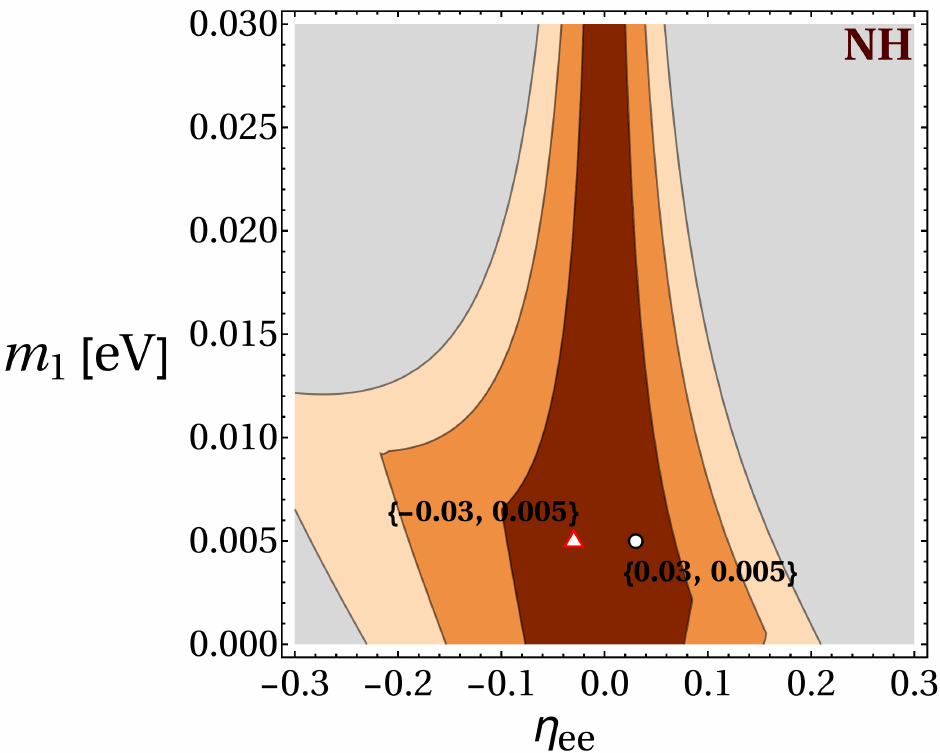}
    \includegraphics[width=0.32\textwidth, height=4.cm]{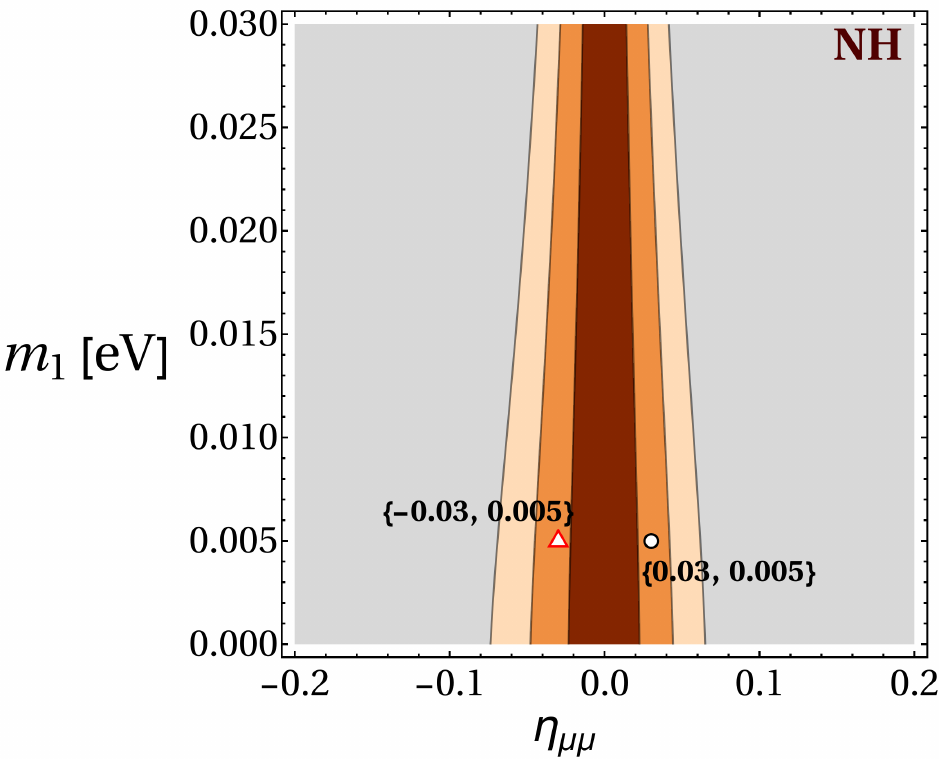}
    \includegraphics[width=0.32\textwidth, height=4.cm]{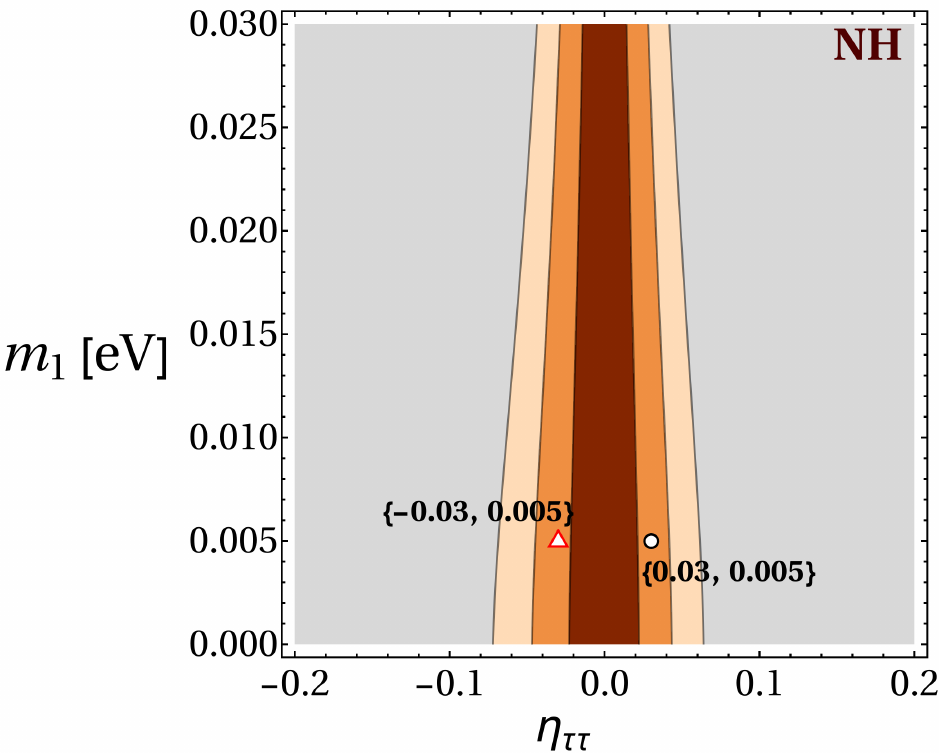}
    \includegraphics[width=0.32\textwidth, height=4.cm]{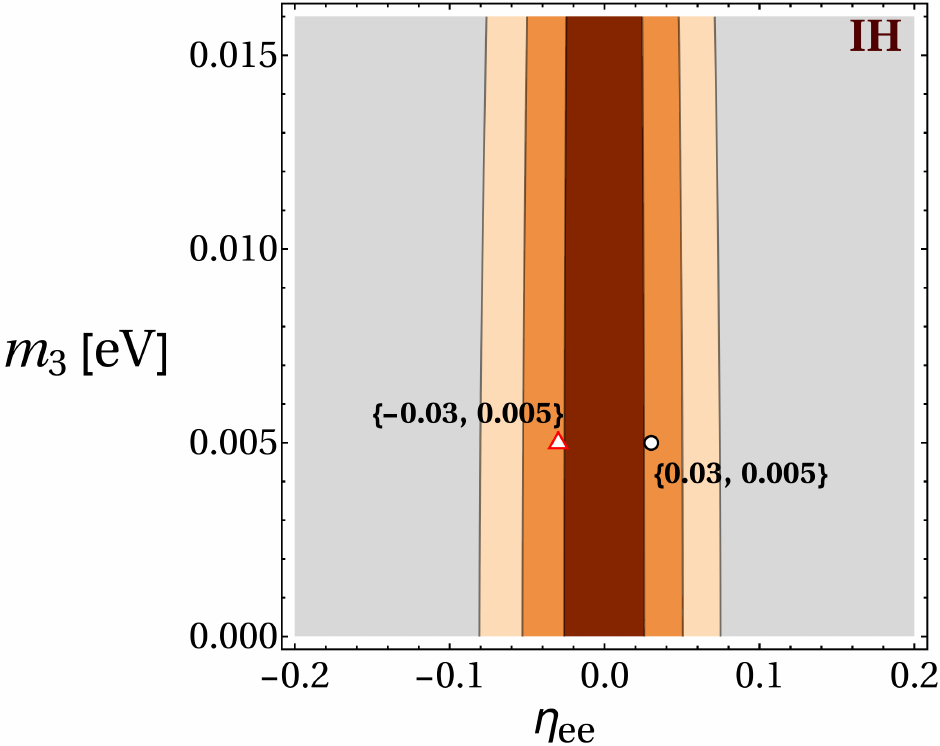}
    \includegraphics[width=0.32\textwidth, height=4.cm]{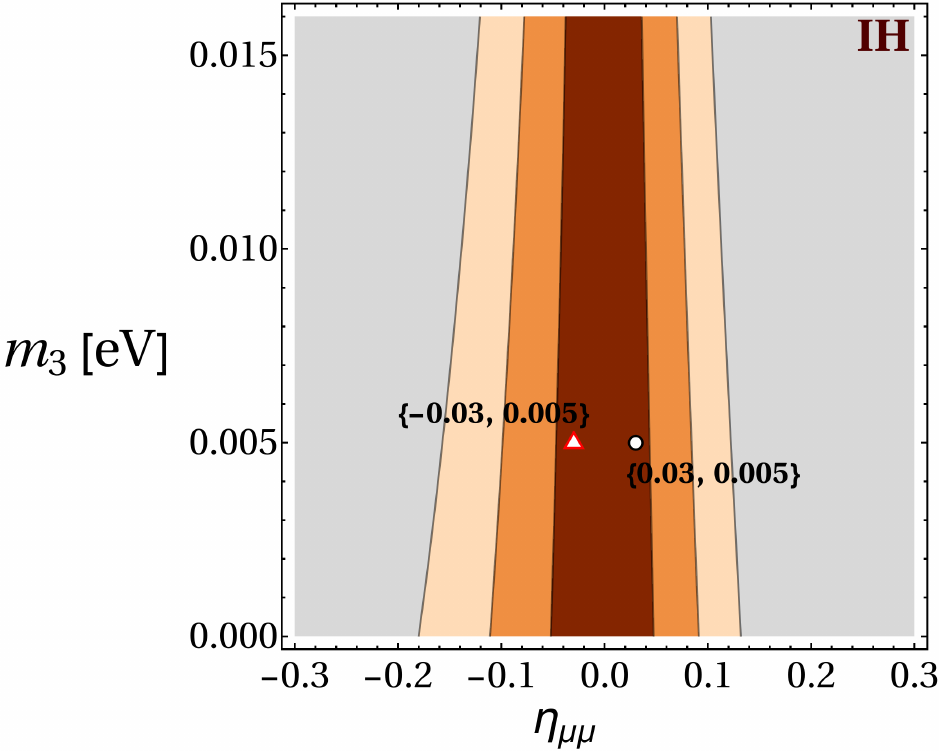}
    \includegraphics[width=0.32\textwidth, height=4.cm]{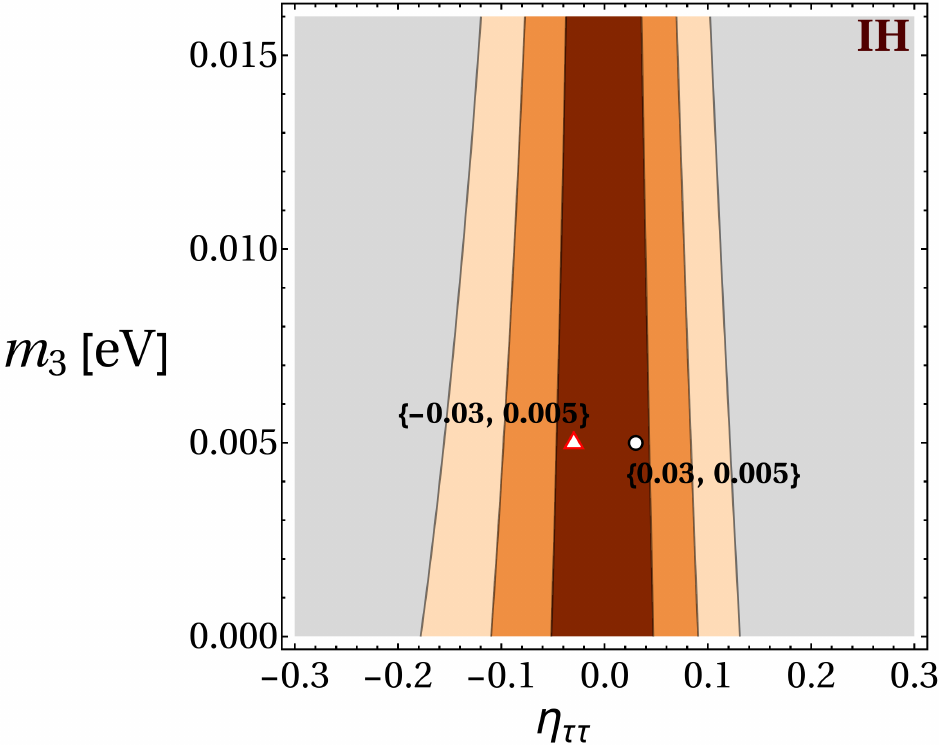}
    \caption{Contour plot for the ratio $\mathcal{R}_{SNSI}$ in the ($\eta_{\alpha\alpha}$ -- $m_\ell$) plane. The top (bottom) panel represents the case of normal (inverted) hierarchy. The left, middle, and right panels correspond to the case of $\eta_{ee}$, $\eta_{\mu\mu}$ and $\eta_{\tau\tau}$. The ($\eta_{\alpha\alpha}$, $m_\ell$) values used for our analysis are highlighted by the markers.}
    \label{fig:SNSI_ratio}
\end{figure*}

\section{} \label{sec:ratio_sec}
In Fig. \ref{fig:SNSI_ratio}, we have explored the parameter space $(\eta_{\alpha\alpha}-m_\ell)$ for the quantity $\mathcal{R}_{SNSI}$. We define a quantity $\mathcal{R}_{SNSI}$ as,
\begin{equation}\label{eq:ratio}
\mathcal{R}_{SNSI} = \Big|\frac{H_{SNSI}\big[i,j\big]}{H_{SI}\big[i,j\big]}\Big|_{max},
\end{equation}
where $H_{SNSI}\big[i,j\big]$ is the $ij^{th}$ element of SNSI contribution to the effective Hamiltonian and $H_{SI}\big[i,j\big]$ is the $ij^{th}$ element of standard Hamiltonian. In addition, we have subtracted the $[1,1]$ element from all the diagonal ratios to remove an overall constant phase. This leaves us with eight ratios out of which we consider the ratio with the maximum value. 

We have explored the parameter space of ($\eta_{\alpha\alpha}$-$m_\ell$) for representative purposes in Fig. \ref{fig:SNSI_ratio}. The top panel represents the normal hierarchy case whereas the bottom panel corresponds to the inverted hierarchy case. The different colors showcase the different contributions of SNSI only at the Hamiltonian level. We have fixed the energy and baseline at L = 1300 km and E = 2.5 GeV considering the DUNE experiment.  Depending on the observed contours, we have restricted our choice of $\eta_{\alpha\alpha}$ such that $\mathcal{R}_{SNSI}$ lies below 0.2. However, it is used only as a representative of the overall SNSI contribution. In our analysis, we have used these values of $\eta_{\alpha\alpha}$ and $m_\ell$ as a case study. The dependence of oscillation probabilities on each SNSI element is also different as shown in equations.

\begin{itemize}
    \item We note that different combinations of the lightest neutrino mass and and $\eta_{\alpha\alpha}$ can lead to similar contributions at the Hamiltonian level. This shows the non-trivial dependence on absolute neutrino masses.
    \item For $\eta_{ee}$, a more narrow band is seen compared to $\eta_{\mu \mu}$ and $\eta_{\tau \tau}$ in case of both hierarchies. However, for IH similar range of $\mathcal{R}_{SNSI}$ is possible for smaller values of $\eta_{ee}$.
    \item For a given $\eta_{\alpha\alpha}$, we observe an increase in value of $\mathcal{R}_{SNSI}$ with an increase in the value of $m_\ell$. We have used this figure only for representative purposes. Based on the relative SNSI contribution, we have chosen certain values of $\eta_{\alpha\alpha}$ such that $\mathcal{R}_{SNSI}$ lie below 0.2.
\end{itemize}
This analysis provides the rationale for selecting the SNSI values for our subsequent exploration.

\section{} \label{app:dcp0}
\begin{figure*}
  \centering
  \includegraphics[height=4.5cm,width=0.32\linewidth]{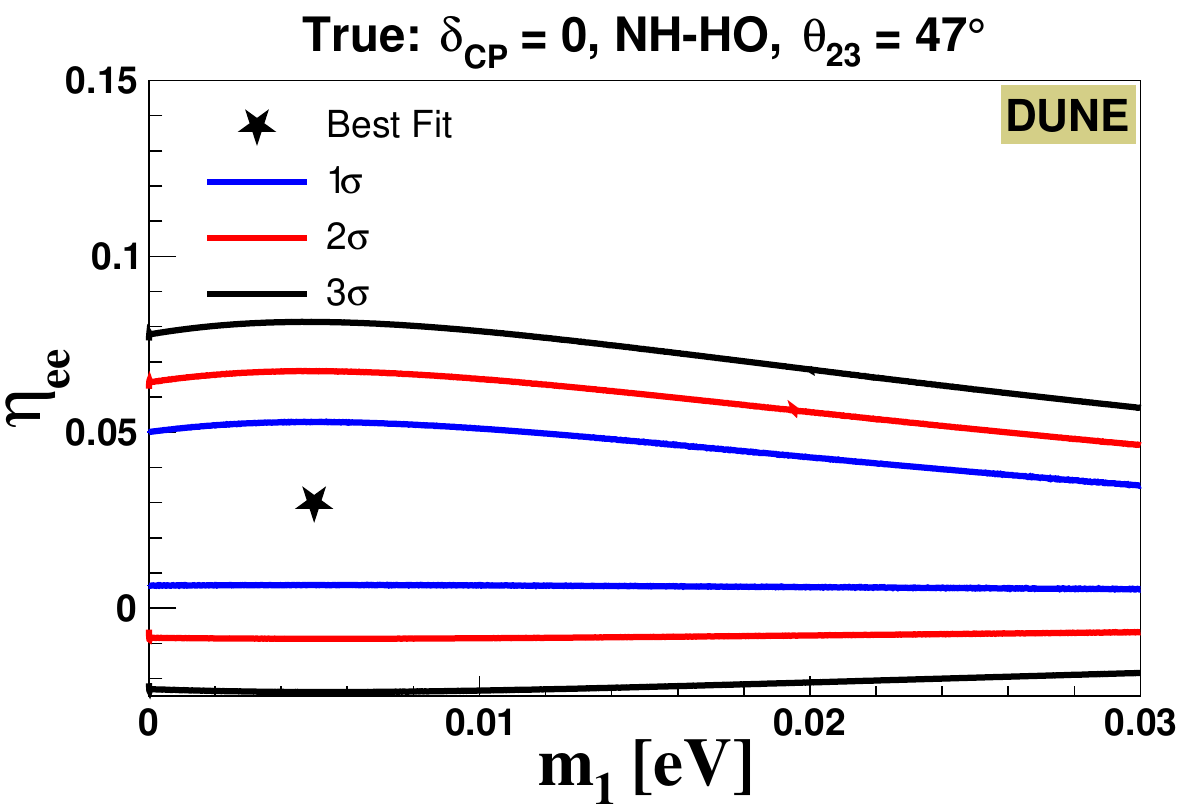}
  \includegraphics[height=4.5cm,width=0.32\linewidth]{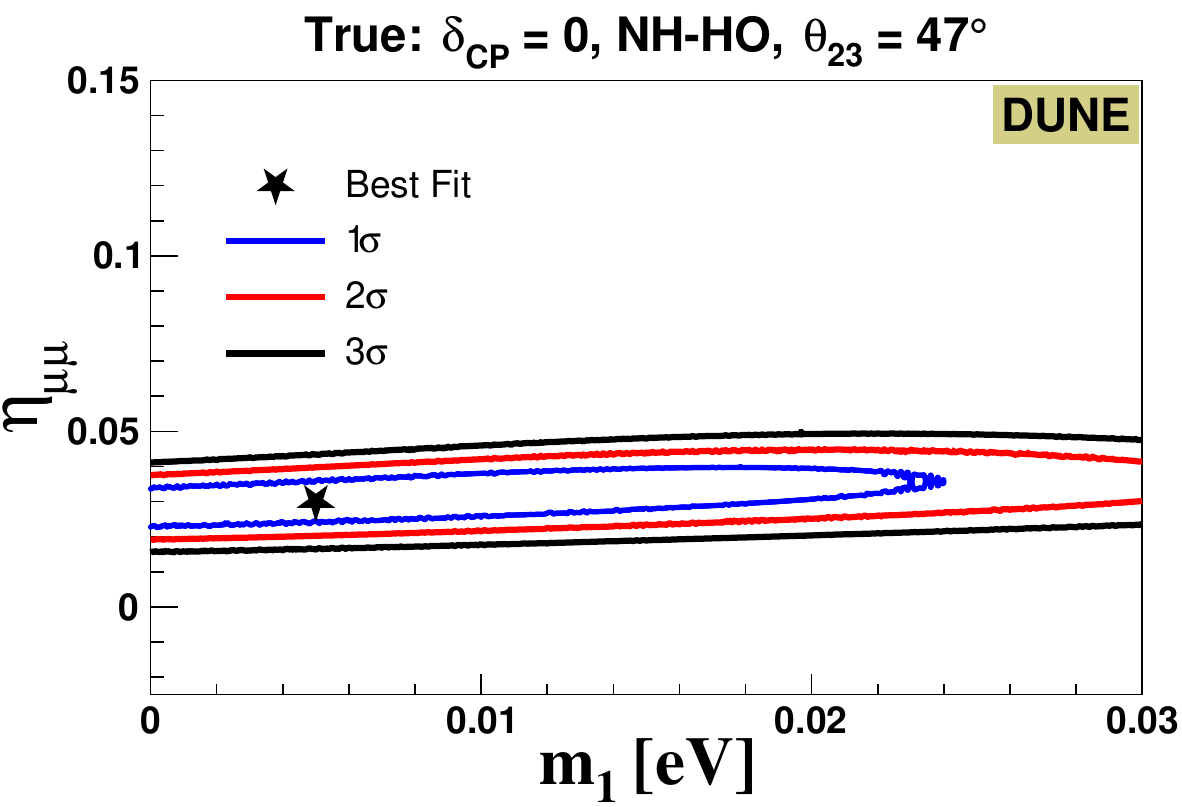}
  \includegraphics[height=4.5cm,width=0.32\linewidth]{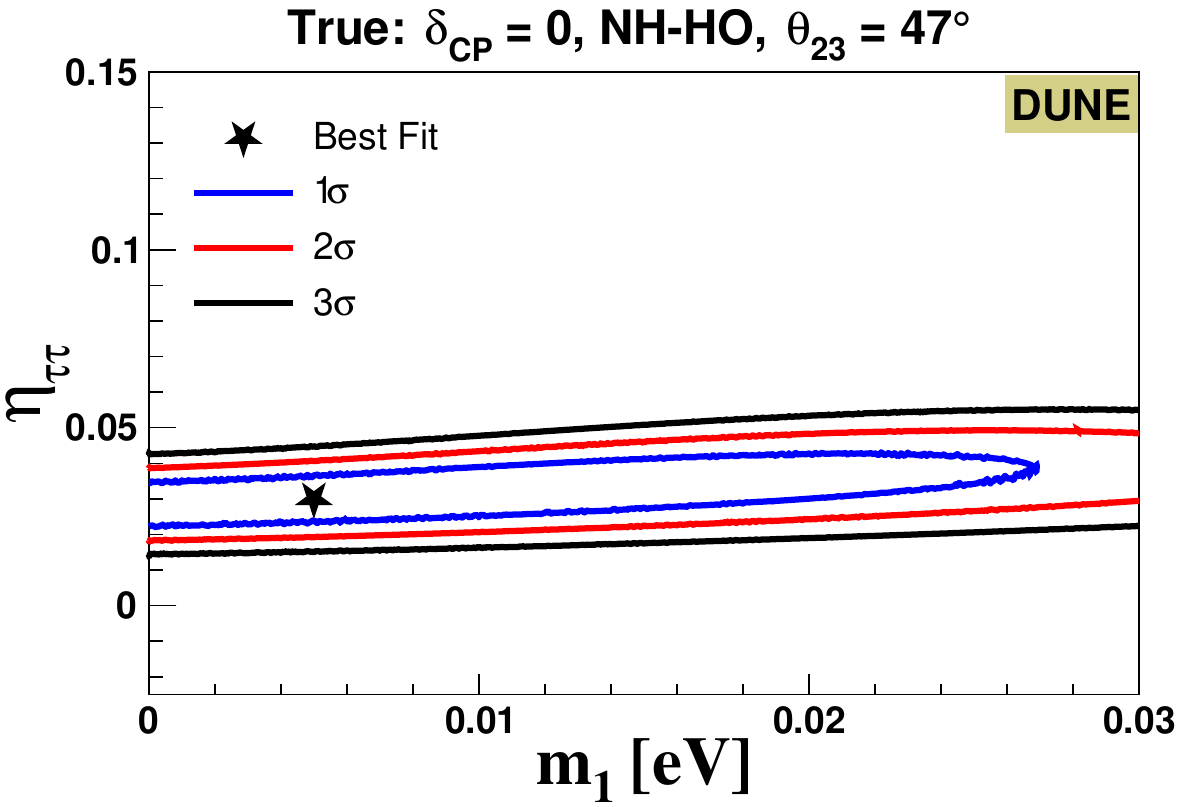}
 \includegraphics[height=4.5cm,width=0.32\linewidth]{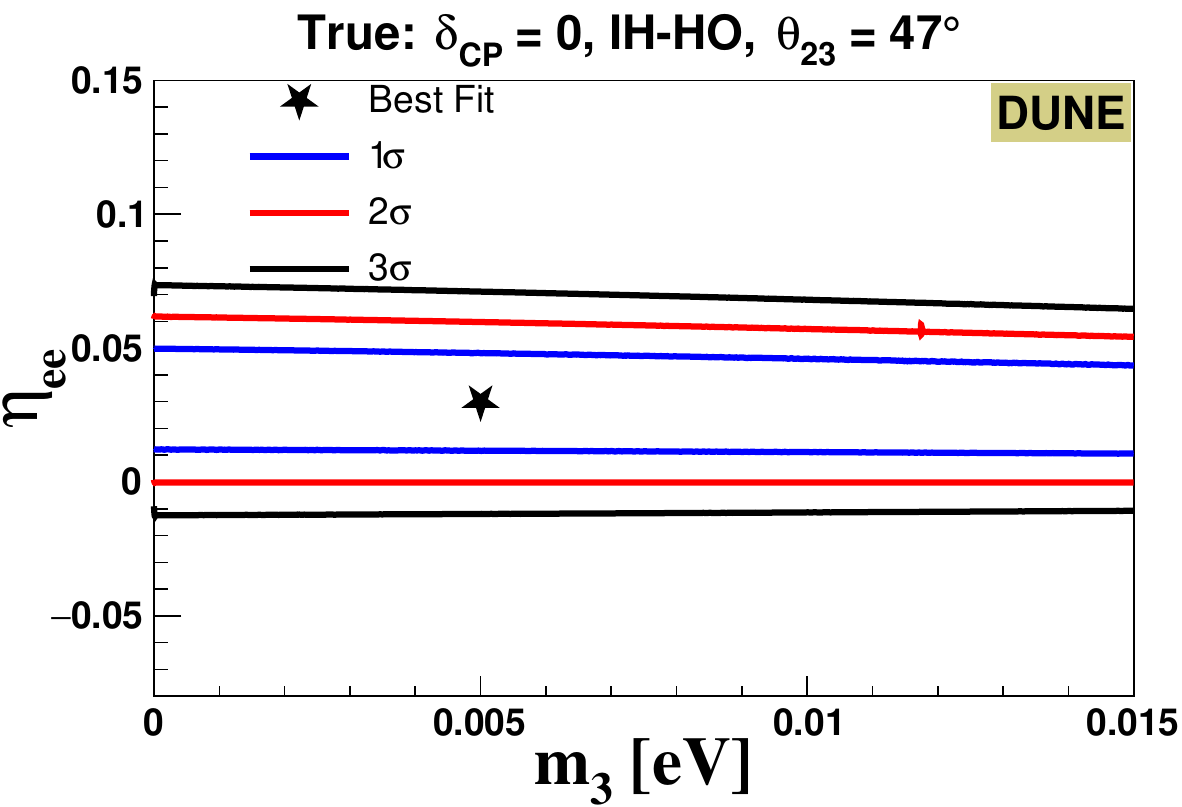}
  \includegraphics[height=4.5cm,width=0.32\linewidth]{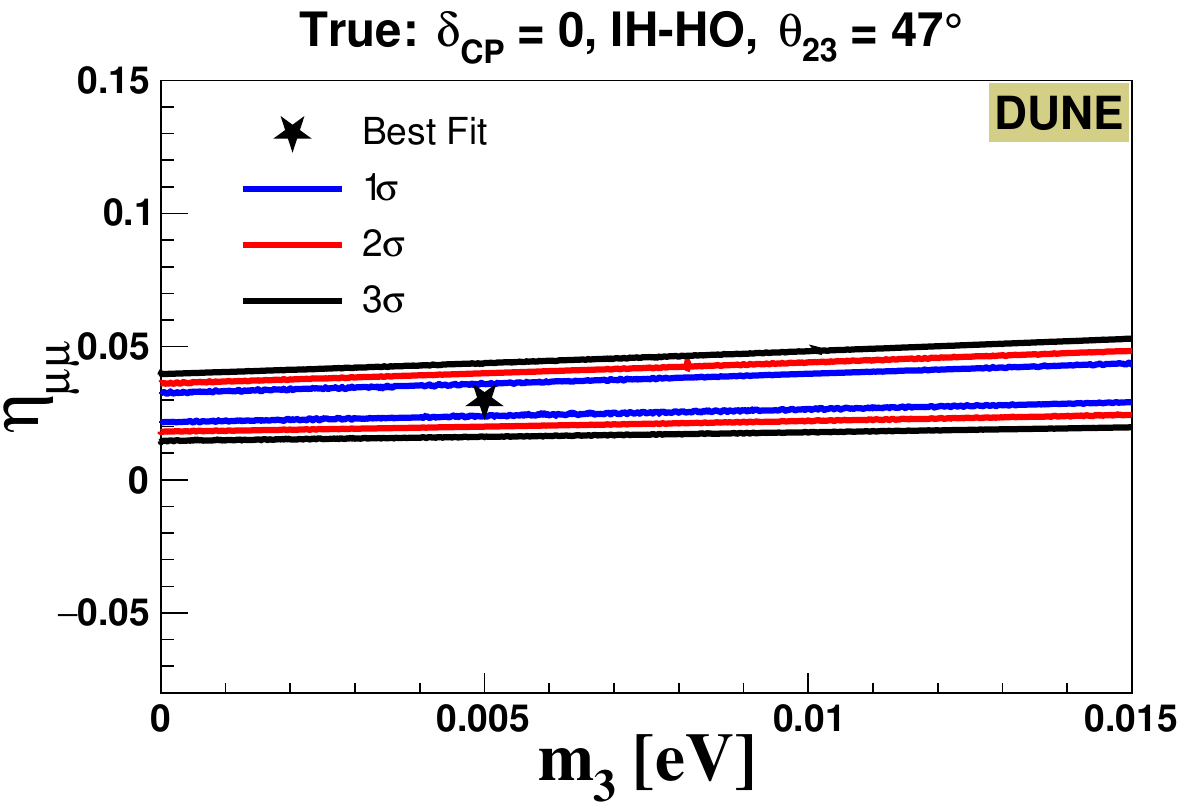}
  \includegraphics[height=4.5cm,width=0.32\linewidth]{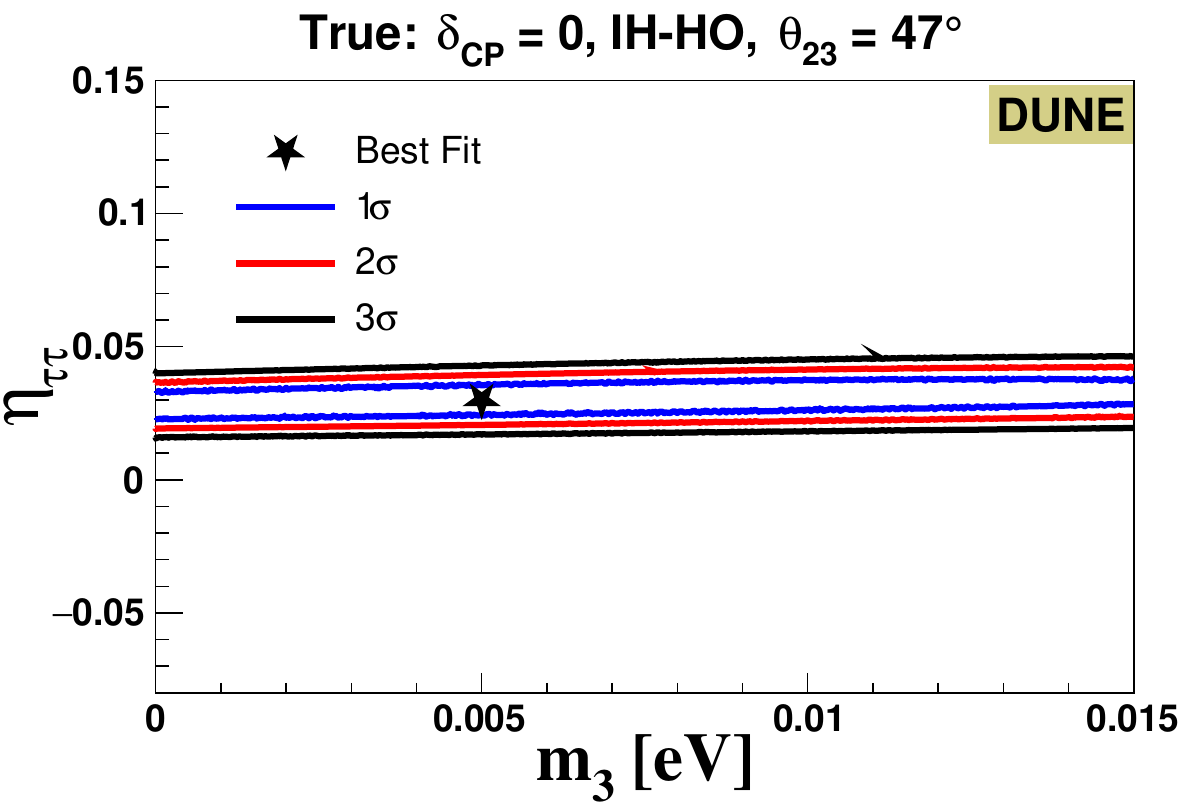}
\caption{\textbf{$\delta_{CP} = 0^{\circ} $, HO:} Allowed region of lightest neutrino mass for normal (top-panel) and inverted (bottom-panel) hierarchy for the diagonal SNSI parameters $\eta_{ee}$ (left-panel), $\eta_{\mu\mu}$ (middle-panel) and $\eta_{\tau\tau}$ (right-panel). The blue, red and magenta lines represent the 1$\sigma$, 2$\sigma$ and 3$\sigma$ confidence levels, respectively. The true value of $m_\ell$ is fixed at $0.005$ eV for NH and IH. The SNSI parameter $\eta_{\alpha\alpha}$ is fixed at 0.03. The best-fit point is represented by the solid black star.}
\label{fig:corr_chi2_3}
\end{figure*}

\begin{figure*}
  \centering
  \includegraphics[height=4.5cm,width=0.32\linewidth]{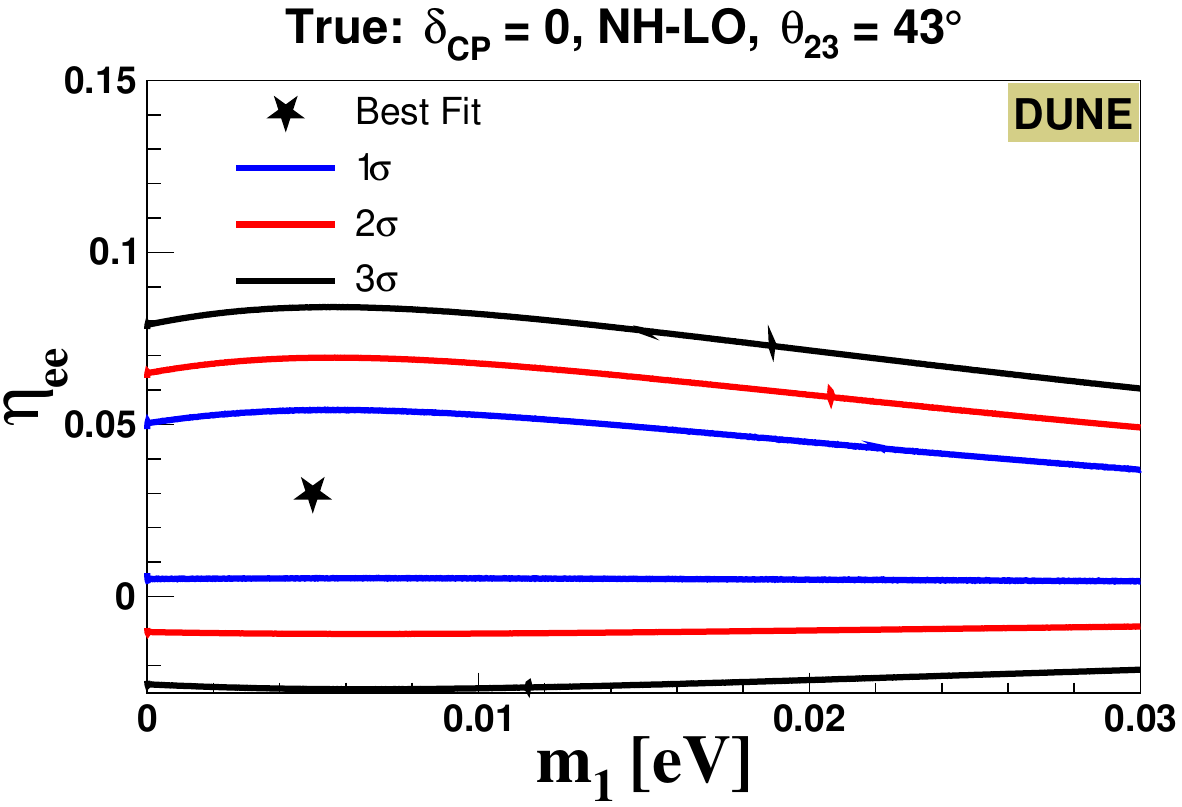}
  \includegraphics[height=4.5cm,width=0.32\linewidth]{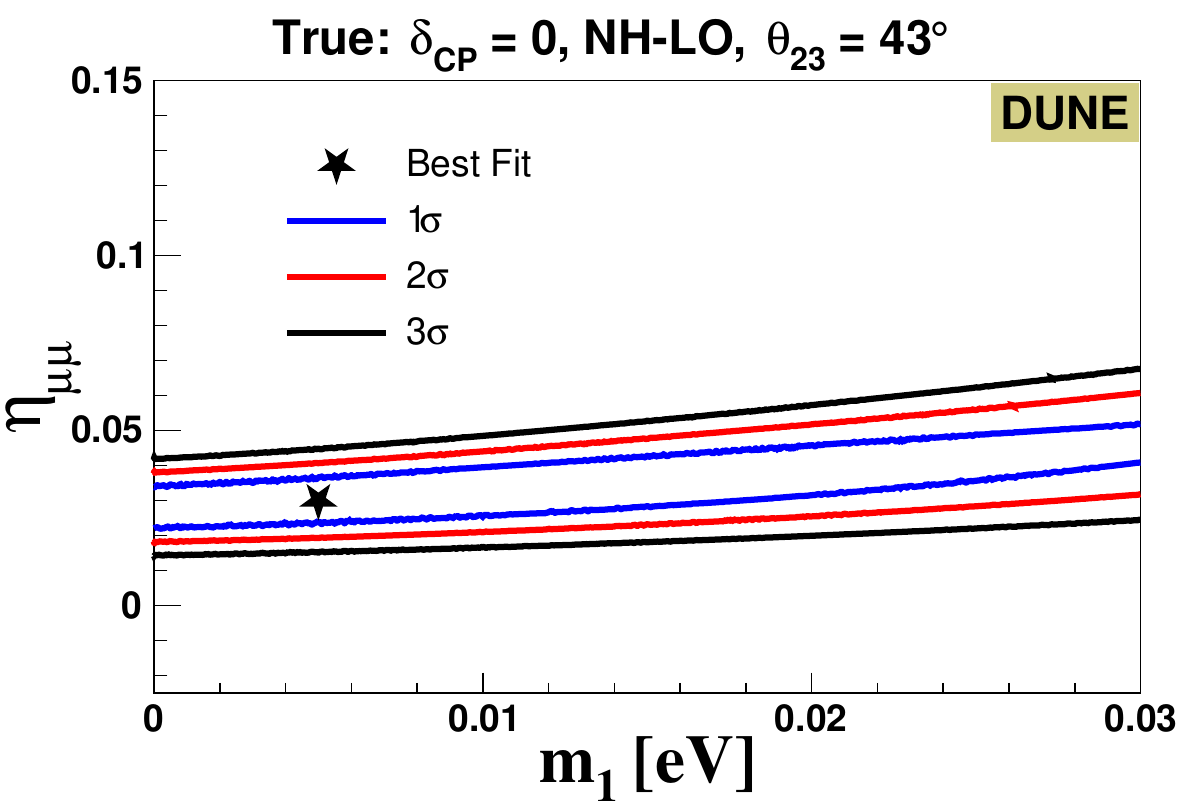}
  \includegraphics[height=4.5cm,width=0.32\linewidth]{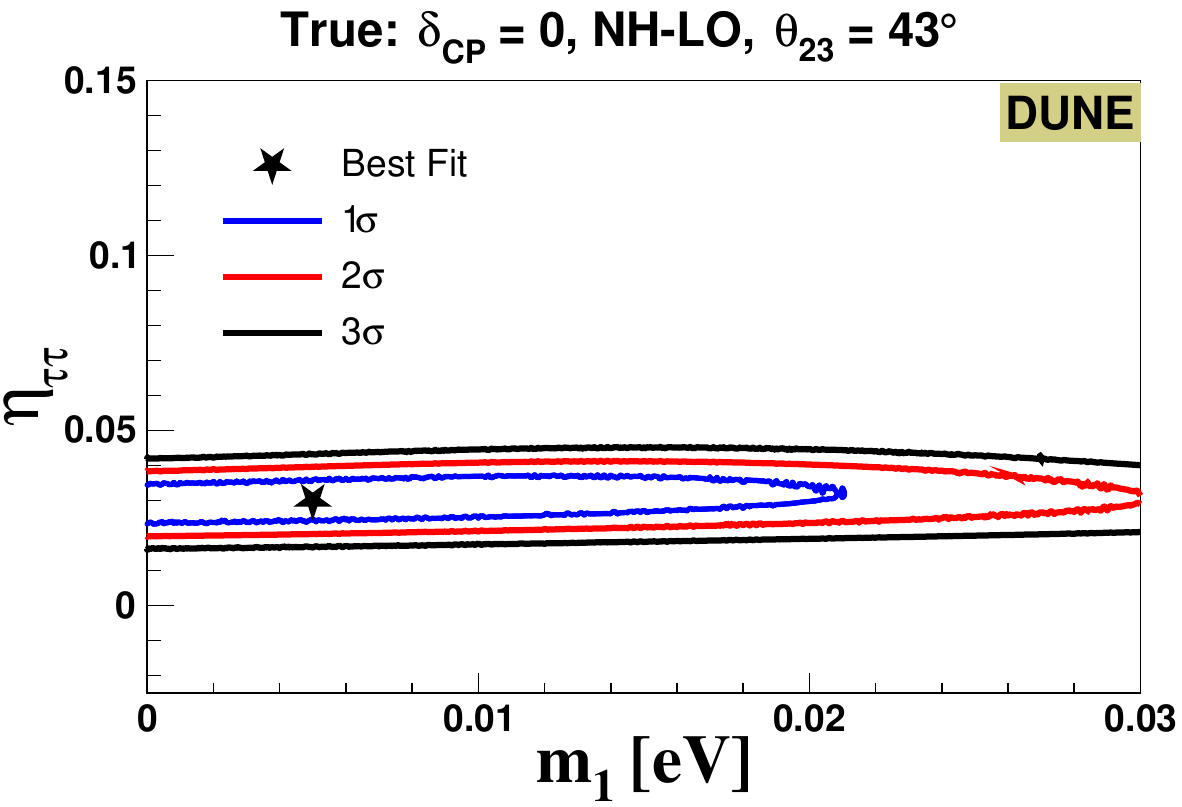}
 \includegraphics[height=4.5cm,width=0.32\linewidth]{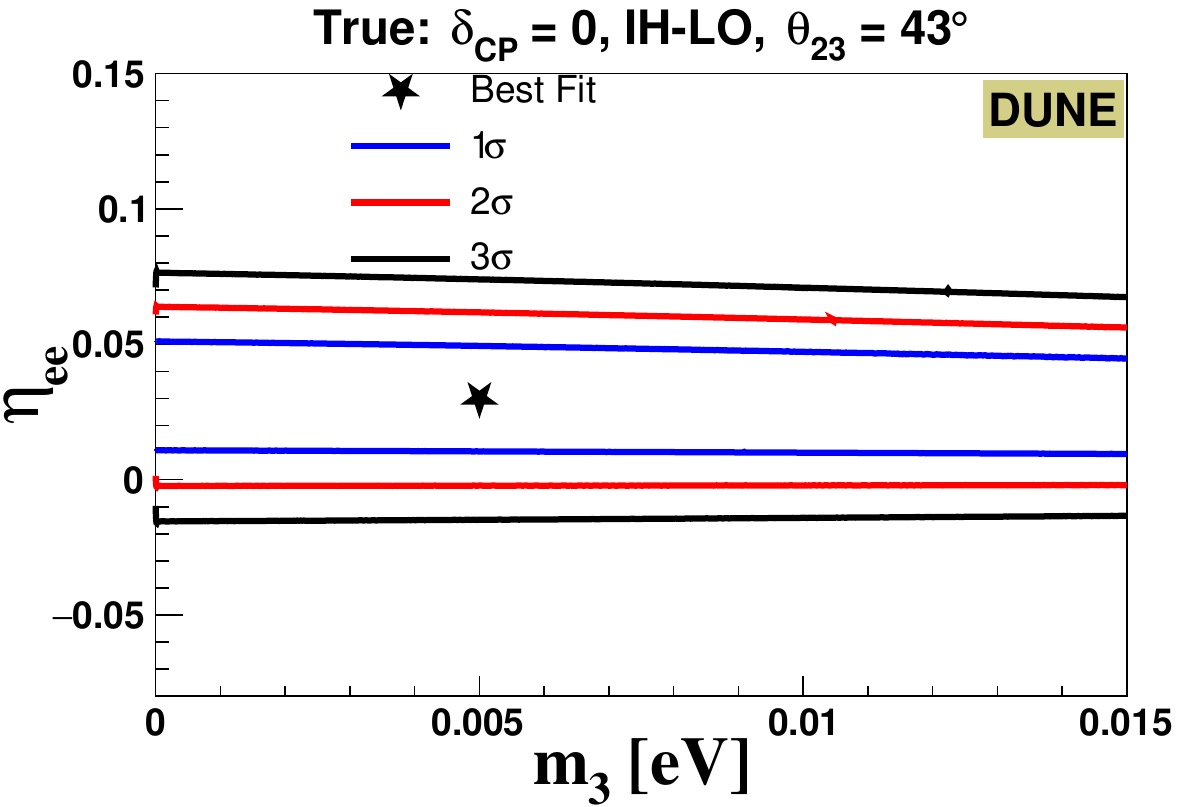}
  \includegraphics[height=4.5cm,width=0.32\linewidth]{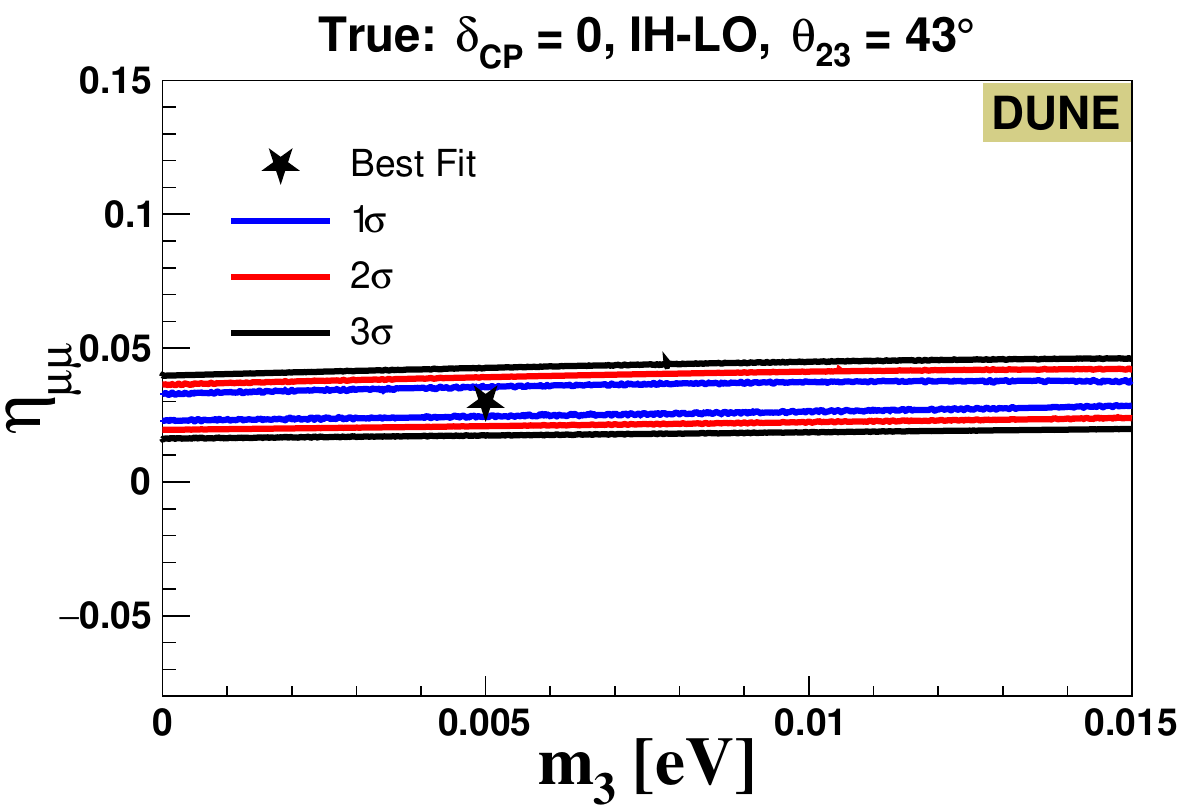}
  \includegraphics[height=4.5cm,width=0.32\linewidth]{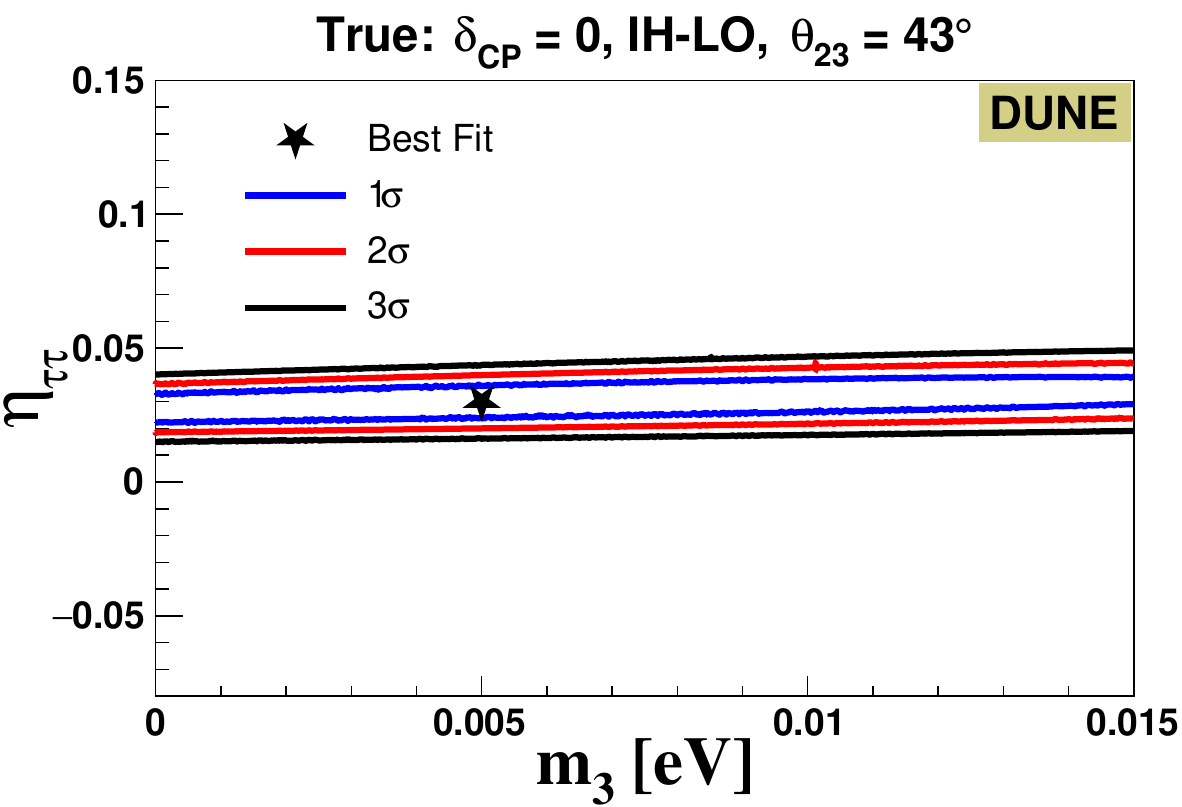}
\caption{\textbf{$\delta_{CP}=0^{\circ}$, LO:} Allowed region of lightest neutrino mass for normal (top-panel) and inverted (bottom-panel) hierarchy for the diagonal SNSI parameters $\eta_{ee}$ (left-panel), $\eta_{\mu\mu}$ (middle-panel) and $\eta_{\tau\tau}$ (right-panel). The blue, red and magenta lines represent the 1$\sigma$, 2$\sigma$ and 3$\sigma$ confidence levels, respectively. The true value of $m_\ell$ is fixed at $0.005$ eV for NH and IH. The SNSI parameter $\eta_{\alpha\alpha}$ is fixed at 0.03. The best-fit point is represented by the solid black star.}
\label{fig:corr_chi2_4}
\end{figure*}

In Fig. \ref{fig:corr_chi2_3} and \ref{fig:corr_chi2_4}, we show the constraints on the lightest neutrino mass in the presence of SNSI for HO and LO respectively at $\delta_{CP}$ = $0^{\circ}$ considering the DUNE experiment. In both the figures the top and bottom panels are for NH and IH respectively. The values of other oscillation parameters are listed in table \ref{tab:param_val}. We set the true values of $(\eta_{\alpha\beta}$, m) at (0.03, 0.005 eV). The blue, red and magenta lines represent the 1$\sigma$, 2$\sigma$ and 3$\sigma$ confidence levels respectively. The best-fit point is represented by a black star in all the plots. We list our observations as follows,

\begin{itemize}
    \item In Fig. \ref{fig:corr_chi2_3}, we observe no significant change in the constraints on $m_\ell$ for all the cases as compared to Fig. \ref{fig:corr_chi2_1}. A similar observation can be made from the comparison of Fig. \ref{fig:corr_chi2_4} and Fig. \ref{fig:corr_chi2_2}.
    \item Our findings demonstrate minimal sensitivity to different values of $\delta_{CP}$. This suggests a degree of robustness in our findings with respect to these oscillation parameters.
\end{itemize}

This analysis demonstrates that varying the CP phase has little effect on the sensitivity of SNSI parameters in constraining neutrino masses. As a result, the ability of the SNSI framework to impose constraints on neutrino masses remains robust, regardless of whether the scenario involves CP conservation or CP violation.
\end{appendices}

\newpage

\bibliography{main}
\end{multicols}

\end{document}